\documentclass[preprint,12pt,authoryear]{elsarticle}

\usepackage[papersize={8.5in,11in}]{geometry}                   

\usepackage[english]{babel}
\usepackage[round]{natbib}                                      
\usepackage{stackrel}
\usepackage{amsmath}
\usepackage{amsfonts}
\usepackage{amssymb}
\usepackage{graphicx}                                           
\usepackage{epstopdf}
\usepackage{caption}
\usepackage{subfigure}                                          
\usepackage[font={footnotesize}]{caption}                       
\usepackage{stmaryrd}
\usepackage{mathtools}
\usepackage{lscape}
\usepackage{color}
\usepackage[colorlinks,citecolor=blue,urlcolor=blue]{hyperref}

\let\oldtabular=\tabular                                        
\def\tabular{\footnotesize\oldtabular}

\date{}                                                         
\makeatletter                                                   
\renewcommand\section{\@startsection {section}{1}{\z@}%
	{-3.5ex \@plus -1ex \@minus -.2ex}%
	{2.3ex \@plus.2ex}%
	{\reset@font\Large\bfseries\centering}}
\makeatother

\DeclareMathOperator{\argmax}{argmax}
\DeclareMathOperator{\BSS}{BSS}
\DeclareMathOperator{\WSS}{WSS}
\DeclareMathOperator{\PLS}{PLS}

\newcounter{desccount}
\newcommand{\descitem}[1]{%
  \item[#1] \refstepcounter{desccount}\label{#1}
}
\newcommand{\descref}[1]{\hyperref[#1]{#1}}

\journal{COMPUTATIONAL STATISTICS \& DATA ANALYSIS}

\begin{document}

\begin{frontmatter}

\title{New developments in Sparse PLS regression.}


\author[adr1,adr2]{J\'er\'emy Magnanensi\corref{cor1}}\ead{magnanensi@math.unistra.fr}
\address[adr1]{Institut de Recherche Math\'ematique Avanc\'ee, UMR 7501, LabEx IRMIA\\
Universit\'e de Strasbourg et CNRS, France}
\address[adr2]{Laboratoire de Biostatistique et Informatique M\'edicale, Facult\'e de M\'edecine, EA3430\\
 Universit\'e de Strasbourg et CNRS, France}
\cortext[cor1]{Corresponding author, 7, Rue Ren\'e Descartes 67084 Strasbourg Cedex, Universit\'e de Strasbourg et CNRS, France} 
\author[adr1]{Myriam Maumy-Bertrand}\ead{mmaumy@math.unistra.fr}
\author[adr2]{Nicolas Meyer}\ead{nmeyer@unistra.fr}
\author[adr1]{Fr\'ed\'eric Bertrand}\ead{fbertrand@math.unistra.fr}

\begin{abstract}
Methods based on partial least squares (PLS) regression, which has recently gained much attention in the analysis of high-dimensional genomic datasets, have been developed since the early 2000s for performing variable selection. Most of these techniques rely on tuning parameters that are often determined by cross-validation (CV) based methods, which raises important stability issues. To overcome this, we have developed a new dynamic bootstrap-based method for significant predictor selection, suitable for both PLS regression and its incorporation into generalized linear models (GPLS). It relies on the establishment of bootstrap confidence intervals, that allows testing of the significance of predictors at preset type I risk $\alpha$, and avoids the use of CV. We have also developed adapted versions of sparse PLS (SPLS) and sparse GPLS regression (SGPLS), using a recently introduced non-parametric bootstrap-based technique for the determination of the numbers of components. We compare their variable selection reliability and stability concerning tuning parameters determination, as well as their predictive ability, using simulated data for PLS and real microarray gene expression data for PLS-logistic classification. We observe that our new dynamic bootstrap-based method has the property of best separating random noise in $\mathbf{y}$ from the relevant information with respect to other methods, leading to better accuracy and predictive abilities, especially for non-negligible noise levels.
Supplementary material is linked to this article.
\end{abstract}





\end{frontmatter}

\section{Introduction}
\label{1}
Partial least squares (PLS) regression, introduced by \citep{wold1983multivariate}, is a well-known dimension-reduction method, notably in chemometrics and  spectrometric modeling \citep{wold2001pls}. In this paper, we focus on the PLS univariate response framework, better known as PLS1. Let $n$ be the number of observations and $p$ the number of covariates. Then, $\mathbf{y}=\left(y_1,\ldots,y_n\right)^T\in\mathbb{R}^{n}$ represents the response vector, with $\left(.\right)^T$ denoting the transpose. The original underlying algorithm, developed to deal with continuous responses, consists of building latent variables $\mathbf{t}_k,\;1\leqslant k\leqslant K$, also called components, as linear combinations of the original predictors $\mathbf{X}=\left(\mathbf{x}_1,\ldots,\mathbf{x}_p\right)\in\mathcal{M}_{n,p}\left(\mathbb{R}\right)$, where $\mathcal{M}_{n,p}\left(\mathbb{R}\right)$ represents the set of matrices of $n$ rows and $p$ columns. Thus,
\begin{equation}
   \mathbf{t}_k=\mathbf{X}_{k-1}\mathbf{w}_k,\:1\leqslant k\leqslant K,
\end{equation}
where $\mathbf{X}_0=\mathbf{X}$, and $\mathbf{X}_{k-1},\;k\geqslant 2$ represents the residual covariate matrix obtained through the ordinary least squares regression (OLSR) of $\mathbf{X}_{k-2}$ on $\mathbf{t}_{k-1}$. Here, $\mathbf{w}_k=\left(w_{1k},\ldots,w_{pk}\right)^T\in\mathbb{R}^p$ is obtained as the solution of the following maximization problem \citep{boulesteix2007partial}:
\begin{align}
\mathbf{w}_k&= \stackrel[\mathbf{w}\in\mathbb{R}^p]{}{\argmax}\left\{\mbox{Cov}^2\left(\mathbf{y}_{k-1},\mathbf{t}_k\right)\right\}\\
&=\stackrel[\mathbf{w}\in\mathbb{R}^p]{}{\argmax}\left\{\mathbf{w}^T\mathbf{X}_{k-1}^T\mathbf{y}_{k-1}\mathbf{y}^T_{k-1}\mathbf{X}_{k-1}\mathbf{w}\right\},
\label{poid}
\end{align}
with the constraint $\left\|\mathbf{w}_k\right\|^2_2=1$, and where $\mathbf{y}_{0}=\mathbf{y}$, and $\mathbf{y}_{k-1},\;k\geqslant 2$ represents the residual vector obtained from the OLSR of $\mathbf{y}_{k-2}$ on $\mathbf{t}_{k-1}$.

The final regression model is thus:
\begin{align}
   \mathbf{y}&=\stackrel[k=1]{K}{\sum{}}c_{k}\mathbf{t}_k+\epsilon \\
						 &=\stackrel[j=1]{p}{\sum{}}\beta_j^{\PLS}\mathbf{x}_j+\epsilon,
				\label{eq3}
\end{align}
with $\mathbf{\epsilon}=\left(\epsilon_1,\ldots,\epsilon_n\right)^{T}$ the $n\, \times \, 1$ error vector and $\left(c_1,\ldots,c_K\right)$ the regression coefficients related to the OLSR of $\mathbf{y}_{k-1}$ on $\mathbf{t}_k$, $\forall k\in\left[\!\left[1,K\right]\!\right]$, also known as $\mathbf{y}$-loadings.
More details are available, notably in \citet{hoskuldsson1988pls} and \citet{Tenenhaus}.

This particular regression technique, based on reductions in the original dimension, avoids matrix inversion and diagonalization, using only deflation. It allows us to deal with high-dimensional datasets efficiently, and notably solves the collinearity problem \citep{wold1984collinearity}. 

With great technological advances in recent decades, PLS regression has been gaining  attention, and has been successfully applied to many domains, notably genomics. Indeed, the development of both microarray and allelotyping techniques result in high-dimensional datasets from which information has to be efficiently extracted. To this end, PLS regression has become a benchmark as an efficient statistical method for prediction, regression and dimension reduction \citep{boulesteix2007partial}. Practically speaking, the observed response related to such studies does commonly not follow a continuous distribution. Frequent goals with gene expression datasets involve classification problems, such as cancer stage prediction, disease relapse, and tumor classification. For such reasons, PLS regression has had to be adapted to take into account discrete responses. This has been an intensive research subject  in recent years, leading globally to two types of adapted PLS regression for classification. The first, studied and developed notably by \citet{nguyen2002multi}, \citet{nguyen2002tumor} and \citet{boulesteix2004pls}, is a two-step method. The first step consists of building  standard PLS components by treating the response as continuous. In the second step, classification methods are run, e.g., logistic discrimination (LD) or quadratic discriminant analysis (QDA). The second type of adapted PLS regression consists of building  PLS components using either an adapted version of or the original iteratively reweighted least squares (IRLS) algorithm, followed by generalized linear regression on these components. This type of method was first introduced by \citet{marx1996iteratively}. Different modifications and improvements, using notably ridge regression \citep{le1992ridge} or Firth's procedure \citep{firth1993bias} to avoid non-convergence and infinite parameter-value estimations, have been developed, notably by \citet{nguyen2004partial}, \citet{ding2012classification}, \citet{fort2005classification} and \citet{bastien2005pls}. In this work, we focus on the second type of adapted PLS regression, referred to from now on as GPLS.\\

As previously mentioned, a feature of datasets of interest is their high-dimensional setting, i.e., $n\ll p$. \citet{chun2010sparse} have shown that the asymptotic consistency of PLS estimators does not hold in this situation, so filtering or predictor selection become necessary in order to obtain consistent parameter estimation. However, all  methods described above proceed to classification using the entire set of predictors. For datasets that frequently contain thousands of predictors, such as microarray ones, a variable filtering pre-processing thus needs to be applied. A commonly used pre-processing method when performing classification uses the $\BSS/\WSS$-statistic:
\begin{equation}
\BSS_j/\WSS_j=\frac{\stackrel[q=1]{Q}\sum{}\stackrel[i:y_i\in G_q]{}\sum{}\left(\hat{\mu}_{jq}-\hat{\mu}_j\right)^2}{\stackrel[q=1]{Q}\sum{}\stackrel[i:y_i\in G_q]{}\sum{}\left(x_{ij}-\hat{\mu}_{jq}\right)^2},
\end{equation}
with $\hat{\mu}_j$ the sample mean of $\mathbf{x}_j$, and $\hat{\mu}_{jq}$ the sample mean of $\mathbf{x}_j$ in class $G_q$ for $q\in\left\{1,\ldots,Q\right\}$. Then, predictors associated with the highest values of this are retained, but no specific rule exists to choose the number of predictors to retain. A Bayesian-based technique, available in the R-package \textit{limma}, has become a common way to deal with this, computing a Bayesian-based p-value for each predictor, therefore allowing users to choose the number of relevant predictors based on a threshold $p$-value \citep{Smyth2004linear}.

This method cannot be considered parsimonious, but rather as a pre-processing stage for exclusion of uninformative covariates. Reliably selecting relevant predictors in PLS regression is of interest for several reasons. Practically speaking, it would allow users to identify the original covariates which are significantly linked to the response, as is done in OLS regression with Student-type tests. Statistically speaking, it would avoid the establishment of over-complex models and ensure consistency of PLS estimators. Several methods for variable selection have already been developed \citep{mehmood2012review}. \citet{lazraq2003selecting} group these techniques into two main categories. The first, model-wise selection, consists of first establishing the PLS model before performing a variable selection. The second, dimension-wise selection, consists of selecting variables on one PLS component at a time. \\

A dimension-wise method, introduced by \citet{chun2010sparse} and called Sparse PLS (SPLS), has become the  benchmark for selecting relevant predictors using PLS methodology. The technique is for continuous responses and consists of simultaneous dimension reduction and variable selection, computing sparse linear combinations of covariates as PLS components. This is achieved by introducing an $L_1$ constraint on the weight vectors $\mathbf{w}_k$, leading to the following formulation of the objective function:

\begin{equation}
\mathbf{w}_k=\stackrel[\mathbf{w}\in\mathbb{R}^p]{}{\argmax}\left\{\mathbf{w}^T\mathbf{X}_{k-1}^T\mathbf{y}_{k-1}\mathbf{y}^T_{k-1}\mathbf{X}_{k-1}\mathbf{w}\right\},\;\text{s.c. }\left\|\mathbf{w}\right\|^2_2=1, \left\|\mathbf{w}\right\|_1<\lambda,
\end{equation}
where $\lambda$ determines the amount of sparsity. More details are available in \citet{chun2010sparse}. Two tuning parameters are thus involved: $\eta\in\left[0,1\right]$ as a rescaled parameter equivalent to $\lambda$, and the number of PLS components $K$, which are determined through CV-based mean squared error (CV MSE). We refer to this technique as SPLS CV in the following. 

\citet{chung2010sparse} have developed an extension of this technique by integrating it into the generalized linear model (GLM) framework, leading it to be able to solve classification problems. They also integrate Firth's procedure, in order to deal with non-convergence issues. Both tuning parameters are selected using CV MSE. We refer to this method as SGPLS CV in the following.\\

A well-known model-wise selection method is the one introduced by \citet{lazraq2003selecting}. It consists of bootstrapping pairs $\left(\mathbf{y}_i,\mathbf{x}_{i\bullet}\right),\;1\leqslant i\leqslant n$, where $\mathbf{x}_{i\bullet}$ represents the $i^{th}$ row of $\mathbf{X}$, before applying PLS regression with a preset number of components $K$ on each bootstrap sample. By performing this method, approximations of distributions related to predictors' coefficients can be achieved. This leads to bootstrap-based confidence intervals (CI) for each predictor, and so opens up the possibility of directly testing their significance with a fixed type I risk $\alpha$. The advantages of this method are twofold. First, by focusing on PLS regression coefficients, it is relevant for both the PLS and GPLS frameworks. Second, it only depends on one tuning parameter $K$, which must be  determined earlier. 

An important related issue should be mentioned. While performing this technique, approximations of distributions are achieved conditionally on the fixed dimension of the extracted subspace. In other words, it approximates the uncertainty of these coefficients, conditionally on the fact that estimations are performed in a $K$-dimensional subspace for each bootstrap sample. The determination of an optimal number of components is crucial for achieving reliable estimations of the regression coefficients \citep{wiklund2007randomization}. Thus, since this determination is specific to the dataset in question, it must be performed for each bootstrap sample, in order to obtain reliable bootstrap-based CI. We have established some theoretical results which confirm this (SI \ref{Theoretical results.}).

Determining tuning parameters by using $q$-fold cross-validation ($q$-CV) based criteria may induce important issues concerning the stability of extracted results \citetext{\citet[p.249]{trevor2001elements}; \citet{BoulesteixCV14}; \citet{magnanensibootcrit}}. Thus, using such criteria for successive choosing of the number of components should be avoided. As mentioned, amongst others, by \citet[p.255]{efron1993introduction} and \citet{kohavi1995study}, bootstrap-based criteria are known to be more stable than CV-based ones. In this context, \citet{magnanensibootcrit} developed a robust bootstrap-based criterion for the determination of the number of PLS components, characterized by a high level of stability and suitable for both the PLS and GPLS regression frameworks. Thus, this criterion opens up the possibility of reliable successive choosing for each bootstrap sample.\\ 

In this article, we introduce a new dynamic bootstrap-based technique for covariate selection suitable for both the PLS and GPLS frameworks. It consists of bootstrapping pairs $\left(\mathbf{y}_i,\mathbf{x}_{i\bullet}\right)$, and  successive extraction of the optimal number of components for each bootstrap sample, using the previously mentioned bootstrap-based criterion. Here, our goal is to better approximate the uncertainty related to regression coefficients by removing the condition of extracting a fixed $K$-dimensional subspace for each bootstrap sample, leading to more reliable CI. This new method both avoids the use of CV, and features the same advantages as those previously mentioned related to the technique introduced by \citet{lazraq2003selecting}. We refer to this new dynamic method as BootYTdyn in the following.

We also succeed in adapting the bootstrap-based criterion introduced by \citet{magnanensibootcrit} to the determination of a unique optimal number of components related to each preset value of $\eta$ in both the SPLS and SGPLS frameworks. Thus, these adapted versions, by reducing the use of CV, improve the reliability of the hyper-parameter tuning. We will refer to  these adapted techniques as SPLS BootYT and SGPLS BootYT, respectively.\\ 

The article is organized as follows. In Section \ref{2}, we introduce our new dynamic bootstrap-based technique, followed by the description of our adaptation of the BootYT stopping criterion to the SPLS and SGPLS frameworks. In Section \ref{3}, we present simulations related to the PLS framework, and summarize the results, depending notably on different noise levels in $\mathbf{y}$. In Section \ref{4}, we treat a real microarray gene expression dataset with a binary response, here the original localization of colon tumors, by benchmarking our new dynamic bootstrap-based approach for GPLS regression. Lastly, we discuss results and  conclusions in Section \ref{5}.

\section{Bootstrap-based approaches for predictor selection}
\label{2}
\subsection{A new dynamic bootstrap-based technique}
\label{21}

As mentioned in Section \ref{1}, the selected number of extracted components is crucial for reliable estimation of $\beta_j^{\PLS},\,1\leqslant j\leqslant p$ \citep{wiklund2007randomization}. We have shown (SI \ref{Theoretical results.}) that selecting an optimal number of components on the original dataset and using it to perform PLS regression on the constructed bootstrap samples, as done by \citet{lazraq2003selecting}, is not correct for the obtention of reliable CI.

In order to take into account these theoretical results, we have developed a new dynamic bootstrap-based approach for variable selection relevant for both the PLS and GPLS frameworks. The approach consists of estimating the optimal number of components for each bootstrap sample created during the algorithm. To obtain consistent results, a robust and resample-stable stopping criterion in component construction has to be used. Let us use $\beta_j$ to mean  $\beta_j^{\PLS}$ in the following, in order to lighten notation. The algorithm for this new  method is  as follows:
\begin{enumerate}
\item Let $D^{ori}$ be the original dataset and $R$ the total number of bootstrap samples $D_r^b,\;r\in\left[\!\left[1,R\right]\!\right]$.
\item $\forall r\in\left[\!\left[1,R\right]\!\right]$:
\begin{itemize}
\item Extract the number of PLS components that is needed for $D_r^b$ using a preset stopping criterion.
\item Compute the estimations $\hat{\beta}_j^r,\;\forall j\in\left[\!\left[1,p\right]\!\right]$, by fitting the relevant PLS or GPLS model.
\end{itemize}
\item $\forall j\in\left[\!\left[1,p\right]\!\right]$, construct a $\left(100\times\left(1-\alpha\right)\right)$\% bilateral $BC_a$ CI for $\beta_j$, noted:
\begin{center}
$\mbox{CI}_j=\left[\mbox{CI}_{j,1},\mbox{CI}_{j,2}\right]$.
\end{center}
\item $\forall j\in\left[\!\left[1,p\right]\!\right]$, \textbf{If} $0\notin \mbox{CI}_j$ \textbf{then} retain $\mathbf{x_j}$ \textbf{else} delete $\mathbf{x_j}$.
\item Obtain the reduced model $\mathcal{M}_{\text{sel}}$ by only integrating the significant predictors, and extracting the number of components $K_{sel}$ determined by the preset stopping criterion.
\end{enumerate}
Note that here, we set $\alpha=0.05$.

\subsection{An adapted bootstrap-based Sparse PLS implementation}
\label{22}

As mentioned by \citet{BoulesteixCV14}, using $q$-CV based methods for tuning parameters  potentially induces problems, notably concerning variability of results due to dependency on the way folds are randomly chosen. However, as detailed in Section \ref{1}, the selection of both tuning parameters involved in the SPLS regression developed by \citet{chun2010sparse} is performed using $q$-CV MSE. Therefore, to improve the reliability of this selection, we adapt the bootstrap-based stopping criterion to this method, which gives the following algorithm:

\begin{enumerate}
\item Let $\left\{\eta_1,\ldots,\eta_s\right\}$ be a set of pre-chosen values for the sparsity parameter and $\left\{k_1,\ldots,k_s\right\}=\left\{1,\ldots,1\right\}$ the set of initial numbers of components for each $\eta_i$. Let $i=1$.
\item Let $c_j^{\eta_i},\;j\in\left[\!\left[1,k_i\right]\!\right]$, be the regression coefficients of $\mathbf{y}$ on $\mathbf{T}_{k_i}=\left(\mathbf{t}_1,\ldots,\mathbf{t}_{k_i}\right)\in\mathcal{M}_{n,k_i}\left(\mathbb{R}\right)$. Obtain $k_i$ $BC_a$ CI for $c_j^{\eta_i},\;j\in\left[\!\left[1,k_i\right]\!\right]$, using the bootstrap-based stopping criterion, noted:
\begin{center}
$\mbox{CI}_{j}^{k_i}=\left[\mbox{CI}_{j,1}^{k_i},\mbox{CI}_{j,2}^{k_i}\right]$.
\end{center}
\item \textbf{If} $\exists j\in\left[\!\left[1,k_i\right]\!\right]\left|0\in\mbox{CI}_{j}^{k_i}\right.$ \textbf{then} $K_{opt}^{\eta_i}=k_i-1$ \textbf{else} $\left\{k_i=k_i+1\text{ and return to step 2}\right\}$.
\item \textbf{While} $i\neq s$ \textbf{then} $\left\{i=i+1\text{ and return to step 2}\right\}$.
\item Return the set of extracted numbers of components $\left\{K_{opt}^{\eta_1},\ldots,K_{opt}^{\eta_s}\right\}$ related to $\left\{\eta_1,\ldots,\eta_s\right\}$.
\item Return the pair $\left(\eta_{opt},K^{\eta_{opt}}_{opt}\right)$ having the lowest CV-based MSE.
\end{enumerate}

Retesting all  components obtained after each increase in $k_i$ is essential since the original predictors involved in the components construction change when $k_i$ increases. \citep{chun2010sparse}. This fact, combined with the aim of retaining orthogonality between components, leads the components themselves to change, so that the significance of each component has to be retested at each step. 

\begin{figure}[h]
		\centering
          \includegraphics[trim = 0cm 0cm 0cm 0cm, clip,scale=0.7]{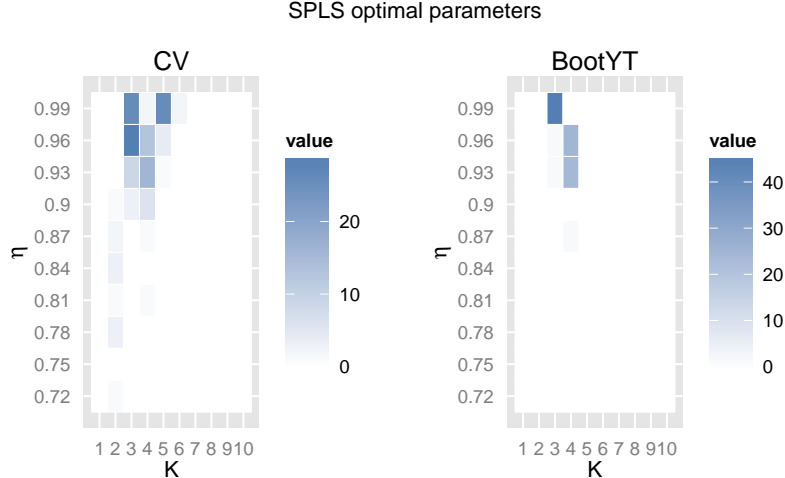}
    \caption{Repartition of 100 selections of $\left(\eta_{opt},K_{opt}\right)$ using the original SPLS approach (left) and the new bootstrap-based implementation (right).}
   \label{fig:1}
\end{figure}

While the original implementation compares $K^{\max}\times s$ models through CV MSE, with $K^{\max}$ the maximal number of components (set by the user), this new bootstrap-based version only focuses on $s$ models, since only one number of components is determined for each preset value of $\eta$. An illustration of the stability improvement obtained by using this new implementation, based on the simulated dataset introduced in Section \ref{321} with $sd\left(\epsilon\right)= 5$, is shown in Fig. \ref{fig:1}.

\section{Simulations studies}
\label{3}

\subsection{Simulations for accuracy comparisons}
\label{31}
These simulations are based on a simulation scheme proposed by \citet[p. 14]{chun2010sparse} and modified in order to study high-dimensional settings. We consider the case where there are less observations than predictors, i.e., $n<p$, and set $n=100$, and $p=200$ or $1000$. Let $q$ be the number of spurious predictors. While \citet[p. 14]{chun2010sparse} only consider a ratio $q/p$ equal to 0.25 and 0.5, both the 0.05 and 0.95 ratio values have been added here. Four independent and identically distributed hidden variables $\mathbf{h}_1,\ldots,\mathbf{h}_4$, following a $\mathcal{N}\left(0,25I_n\right)$ distribution, were computed. Then, columns of the covariate matrix $\mathbf{X}$ were generated by $\mathbf{x}_j=\mathbf{h}_l+\epsilon_j$ for $p_{l-1}+1\leqslant j\leqslant p_l$, where $l=1,\ldots,4$, $\left(p_0,\ldots,p_4\right)=\left(0,\left(p-q\right)/2,p-q,p-r,p\right)$, $r=5$ when $p=200$ and $r=10$ when $p=1000$, and $\epsilon_1,\ldots,\epsilon_p$ are drawn independently from a $\mathcal{N}\left(0,0.1I_n\right)$. Also, $\mathbf{y}$ is generated by $3\mathbf{h}_1-4\mathbf{h}_2+f$, where $f$ is normally distributed with mean 0 and variance such that the signal-to-noise ratio (SNR) equals 10. 

Using this simulation scheme, accuracy of the SPLS technique using 10 fold-CV for  selecting tuning parameters (SPLS CV), and our new dynamic bootstrap method combined with the bootstrap-based stopping criterion (BootYTdyn), is compared. In order to do so, for each parameter setting, 50 selections of the sparse support related to both methods were established. Lastly, mean accuracy values  over the 50 trials were calculated. Results are summarized in Table \ref{tab.1a}.

\begin{table}[h]
	\caption{Mean accuracy values (SNR). \label{tab.1a}}
	  \centering
		\begin{tabular}{lrrrrrrrrr}
			\hline\hline
   $p$ & \multicolumn{4}{c}{200} & & \multicolumn{4}{c}{1000}\\
	\cline{2-5}\cline{7-10}
	 $q/p$  	& \multicolumn{1}{c}{0.05} & \multicolumn{1}{c}{0.25} & \multicolumn{1}{c}{0.5} & \multicolumn{1}{c}{0.95} & & \multicolumn{1}{c}{0.05} & \multicolumn{1}{c}{0.25} & \multicolumn{1}{c}{0.5} & \multicolumn{1}{c}{0.95}\\
		\hline
  SPLS CV & 0.986 & 0.961 & 0.849 & 0.591 & & 0.998 & 0.997 & 0.989 & 0.963 \\
	BootYTdyn & 1.000 & 0.867 & 0.805 & 0.982 & & 0.967 & 0.827 & 0.893 & 0.985 \\
			\hline
			\end{tabular}
\end{table}

Based on these results,  SPLS CV gives better accuracy than  BootYTdyn for ratio values $q/p$ that are not close to 0 or 1. While both give good performance when the ratio is close to 0,  i.e., when a major part of predictors are significant,  BootYTdyn outperforms SPLS CV when only a small proportion of predictors are significant. 

Nevertheless, in this simulation set-up, covariates are collected into four groups. While within-group correlations between covariates are close to one,  between-group ones are close to zero. This unrealistic situation makes irrelevant the determination of an optimal support, and seems more appropriate to selecting the number of components. As an illustration, 50 additional samples in the $p=1000$ and $q/p=0.5$ case were simulated. We then calculated the predictive MSE (PMSE) based on four different supports $S_1,S_2,S_3,S_4$, where $S_1=\left\{\mathbf{x}_j,1\leqslant j\leqslant p\right\}$, $S_2=\left\{\mathbf{x}_j,1\leqslant j\leqslant \left(p-q\right)\right\}$, $S_3=\left\{\mathbf{x}_1,\mathbf{x}_{251}\right\}$ and $S_4=\left\{\mathbf{x}_1,\mathbf{x}_{251}\right\}\cup\left\{\mathbf{x}_j,\left(p-q\right)+1\leqslant j\leqslant p\right\}$. Results are summarized in Table \ref{tab.2a}.

\begin{table}[ht]
	\caption{PMSE values for different supports.  \label{tab.2a}}
  \centering
		\begin{tabular}{lrrrr}
			\hline\hline
	 	& \multicolumn{1}{c}{$S_1$} & \multicolumn{1}{c}{$S_2$} & \multicolumn{1}{c}{$S_3$} & \multicolumn{1}{c}{$S_4$}\\
		\hline
  K=1 & 65.383 & 62.702 & 63.988 & 856.779  \\
	K=2 & 63.957 & 64.443 & 65.695 & 209.873  \\
	K=3 & 65.745 & 76.197 & NA & 61.801 \\
			\hline
		\end{tabular}
\end{table}

In the light of these observations, the aim of this simulation scheme  is instead the extraction if an optimal number of components rather than an optimal support. We thus decided to use a real dataset as covariate matrix for a more general and relevant comparison.

\subsection{Simulations for global comparison}
\label{32}
\subsubsection{Dataset simulations}
\label{321}

In this study, we used a real microarray gene expression dataset, which was created using fresh-frozen primary tumors samples, from a multi-center cohort, with stage I to IV  colon cancer. 566 samples fulfilled RNA quality requirement, and constituted our database. These samples were split into a 443 sample discovery set and a 123 sample test set, well balanced for the main anatomo-clinical characteristics. This database has already been studied by \citet{marisa2013gene} and more details on it are available in their article. 

In order to reduce computational time, a preliminary selection of 100 predictors was performed. Based on the original localization of the tumors as response vector, and the full 566 samples, the 100 most differentially expressed probe sets were extracted. As mentioned in Section \ref{1}, this pre-processing is based on a Bayesian technique and gives us our benchmark predictors matrix.\\

Then, based on correlation values, four positively-correlated predictors were selected to form the set of significant covariates (SI \ref{Predictors' correlations properties.}). To this end, let $\mathbf{X}_{sel}=\left(\mathbf{x}_1,\mathbf{x}_{12},\mathbf{x}_{15},\mathbf{x}_{59}\right)\in\mathcal{M}_{n,4}\left(\mathbb{R}\right)$ be the matrix composed of these predictors, so that $\mathbf{y}$ is simulated as follows:

\begin{equation}
 \mathbf{y}=\mathbf{X}_{sel}\mathbf{\beta}+\epsilon,
\label{eq:1}
\end{equation}
with $\beta=\left(3.559,2.071,1.440,1.770\right)^T$, $\mathbb{E}\left(\epsilon\right)=0$ and Var$\left(\epsilon\right)=\sigma^2I_n$.

We performed simulations for six distinct levels of random noise standard deviation in order to investigate the performance of the different methods. Both these standard deviations and their related SNR are shown in Table \ref{tab.2}.

\begin{table}[ht]
	\caption{Noise standard deviation (SNR).  \label{tab.2}}
   \centering
		\begin{tabular}{rrrrrr}
			\hline\hline
     \multicolumn{1}{c}{dataset 1} & \multicolumn{1}{c}{dataset 2} & \multicolumn{1}{c}{dataset 3} & \multicolumn{1}{c}{dataset 4} & \multicolumn{1}{c}{dataset 5} & \multicolumn{1}{c}{dataset 6}\\
		\hline
    0.5 (810.603) & 1 (202.651) & 3 (22.517) & 4 (12.666) & 5 (8.106) & 6.366 (5.000)\\
			\hline
	\end{tabular}
\end{table}

\subsubsection{Benchmarked methods}
\label{322}

Using these simulated datasets, eight methods were analyzed and compared.
\begin{enumerate}
\item $\mathbf{Q^2}$. The original bootstrap-based method, introduced by \citet{lazraq2003selecting}, combined with the 10-fold CV-based $Q^2$ criterion \citep[p. 83]{Tenenhaus} for pre-selecting the number of components.
\item \textbf{BIC}. The bootstrap-based method, introduced by \citet{lazraq2003selecting}, combined with the corrected BIC using  estimated degrees of freedom (DoF) \citep{kramer2011degrees} for pre-selecting the number of components.
\item \textbf{BootYT}. The bootstrap-based method, introduced by \citet{lazraq2003selecting}, combined with the bootstrap-based criterion \citep{magnanensibootcrit} for previously selecting the number of components.
\item \textbf{BICdyn}. Our new dynamic bootstrap-based method combined with the corrected BIC criterion for successive selections of number of components.
\item \textbf{BootYTdyn}. Our new dynamic bootstrap-based method combined with the bootstrap-based criterion for successive selections of number of components.
\item \textbf{SPLS CV}. The original SPLS method using 10-fold CV for tuning parameter determination \citep{chun2010sparse}.
\item \textbf{SPLS BootYT}. The new adapted SPLS version using the bootstrap-based criterion for component selection.
\item \textbf{Lasso}. Lasso regression, included as a benchmark \citep{efron2004least}.
\end{enumerate}

\subsubsection{Simulation scheme and notation}
\label{323}

In order to perform reliable comparisons between the eight methods, each type of trial was performed one hundred times. Numbers of components, sparse supports and sparse tuning parameters are the main examples of these. Results linked to the highest occurrence rates are then chosen for method comparison. All bootstrap-based techniques were performed with $R=1000$ bootstrap samples, and each related CI was constructed with type I risk $\alpha=0.05$. 

The global comparison has two main parts. First, in order to compare accuracy and stability related to each technique, we focus both on different supports, and models extracted by the different variable selection methods. Indeed, in the PLS framework, a specific model results from both a set of predictors and a specific number of components. Due to the sparsity parameter in SPLS approaches, the same support can be extracted, but with a different number of components  leading to different models. Lasso regression can also extract the same support for several different sparsity parameters, leading to different estimations of model coefficients. Therefore, for clarity, the following notation, related to each specific variable selection technique, is introduced.
\begin{itemize}
\item $\left\{\mathcal{S}_1,\ldots,\mathcal{S}_{\Gamma_1}\right\}$, the set of extracted supports.
\item $\left\{\mathcal{M}_1,\ldots,\mathcal{M}_{\Gamma_2}\right\}$, the set of  fitted models.
\item $\mathcal{S}_{sel}$, the selected support, i.e., the one that appears the most often. 
\item $\mathcal{M}_{sel}$, the selected model, i.e., the one that appears the most often. 
\item $\%\mathcal{S}_{sel}$, rate of occurrence of the selected support.
\item $\%\mathcal{M}_{sel}$, rate of occurrence of the selected model.
\item $K_{sel}$, the number of components related to the selected model.
\end{itemize}

Second, in order to compare the predictive ability of models, 10-fold CV MSE, related to each selected sparse model through PLS regression, were computed one hundred times. The test set was also used in order to confirm results obtained by CV.

Note that, concerning the dynamic BIC-based method for $sd\left(\epsilon\right)= 0.5$, only 97 trials performed well. Lastly, due to equality of occurrence rates between the two most-represented pairs of tuning parameters, results for SPLS CV and sd$\left(\epsilon\right)= 5$ come instead from 150 trials.

\subsubsection{Stability and accuracy results}
\label{324}

Both the mean accuracy values over trials in each case, and stability results based on extracted supports, are given in Tables \ref{tab.3}, \ref{tab.4} and \ref{tab.5}. The numbers of components used for the original bootstrap-based approach \citep{lazraq2003selecting} are summarized in SI \ref{Numbers of extracted components.}.

\begin{table}[ht]
	\caption{Mean accuracy values.  \label{tab.3}}
  \centering
		\begin{tabular}{lrrrrrrrr}
			\hline\hline
       & \multicolumn{1}{c}{$Q^2$} & \multicolumn{1}{c}{BIC} & \multicolumn{1}{c}{BICdyn} & \multicolumn{1}{c}{BootYT} & \multicolumn{1}{c}{BootYTdyn} & \multicolumn{1}{c}{SPLS CV} & \multicolumn{1}{c}{SPLS BootYT} & \multicolumn{1}{c}{Lasso}\\
		\hline
  sd$\left(\epsilon\right)= 0.5$ & 0.9331 & 0.9710 & 0.9882 & 0.9587 & 0.9718 & 0.9914 & 0.9960 & 0.9572\\
	sd$\left(\epsilon\right)= 1$ & 0.9370 & 0.9503 & 0.9575 & 0.9557 & 0.9781 & 0.9915 & 1.0000 & 0.9689\\ 
	sd$\left(\epsilon\right)= 3$ & 0.8730 & 0.9353 & 0.9576 & 0.9614 & 0.9837 & 0.9821 & 0.9799 & 0.9741\\ 
	sd$\left(\epsilon\right)= 4$ & 0.8004 & 0.9289 & 0.9397 & 0.9692 & 0.9928 & 0.9771 & 0.9799 & 0.9686\\ 
	sd$\left(\epsilon\right)= 5$ & 0.8327 & 0.9176 & 0.9444 & 0.9557 & 0.9876 & 0.9790 & 0.9841 & 0.9676\\ 
	sd$\left(\epsilon\right)= 6.366$ & 0.8970 & 0.8755 & 0.9347 & 0.9625 & 0.9820 & 0.9745 & 0.9714 & 0.9731\\ 
			\hline
		\end{tabular}
\end{table}

\begin{table}[ht]
	\caption{Number $\Gamma_1$ of different extracted supports $\left(\%\mathcal{S}_{sel}\right)$.  \label{tab.4}}
  \centering
		\begin{tabular}{lrrrrrrrr}
			\hline\hline
       & \multicolumn{1}{c}{$Q^2$} & \multicolumn{1}{c}{BIC} & \multicolumn{1}{c}{BICdyn} & \multicolumn{1}{c}{BootYT} & \multicolumn{1}{c}{BootYTdyn} & \multicolumn{1}{c}{SPLS CV} & \multicolumn{1}{c}{SPLS BootYT} & \multicolumn{1}{c}{Lasso}\\
		\hline
  sd$\left(\epsilon\right)= 0.5$ & 20 (17) & 23 (10) & 6 (73.2) & 11 (30) & 16 (35) & 5 (56) & 2 (90) & 4 (48)\\
	sd$\left(\epsilon\right)= 1$ & 18 (30) & 8 (57) & 7 (38) & 11 (26) & 17 (23) & 6 (53) & 1 (100) & 5 (49)\\ 
	sd$\left(\epsilon\right)= 3$ & 41 (19) & 16 (47) & 14 (39) & 26 (16) & 6 (44) & 5 (34) & 4 (90) & 4 (58)\\ 
	sd$\left(\epsilon\right)= 4$ & 96 (2) & 6 (58) & 11 (26) & 18 (30) & 6 (48) & 12 (48) & 4 (67) & 4 (69)\\ 
	sd$\left(\epsilon\right)= 5$ & 88 (3) & 11 (48) & 17 (33) & 12 (22) & 6 (54) & 11 (25.33) & 4 (45) & 3 (64)\\ 
	sd$\left(\epsilon\right)= 6.366$ & 47 (10) & 25 (24) & 9 (57) & 10 (39) & 5 (38) & 10 (18) & 4 (46) & 3 (64)\\ 
			\hline
		\end{tabular}
\end{table}

\begin{table}[ht]
	\caption{Number of predictors in $\mathcal{S}_{sel}$ $\left(K_{sel}\right)$.  \label{tab.5}}
  \centering
		\begin{tabular}{lrrrrrrrr}
			\hline\hline
       & \multicolumn{1}{c}{$Q^2$} & \multicolumn{1}{c}{BIC} & \multicolumn{1}{c}{BICdyn} & \multicolumn{1}{c}{BootYT} & \multicolumn{1}{c}{BootYTdyn} & \multicolumn{1}{c}{SPLS CV} & \multicolumn{1}{c}{SPLS BootYT} & \multicolumn{1}{c}{Lasso}\\
		\hline
  sd$\left(\epsilon\right)= 0.5$ & 11 (5) & 6 (4) & 5 (5) & 8 (5) & 7 (6) & 4 (4) & 4 (4) & 4 (4)\\
	sd$\left(\epsilon\right)= 1$ & 10 (4) & 9 (5) & 9 (5) & 9 (5) & 6 (4) & 4 (4) & 4 (4) & 5 (3)\\ 
	sd$\left(\epsilon\right)= 3$ & 16 (4) & 11 (6) & 8 (6) & 7 (4) & 6 (4) & 8 (4) & 6 (4) & 7 (4)\\ 
	sd$\left(\epsilon\right)= 4$ & 24 (3) & 11 (5) & 10 (5) & 6 (4) & 5 (3) & 6 (4) & 6 (4) & 7 (4)\\ 
	sd$\left(\epsilon\right)= 5$ & 21 (3) & 12 (5) & 10 (5) & 9 (4) & 3 (3)  & 5 (3) & 3 (3) & 7 (4)\\ 
	sd$\left(\epsilon\right)= 6.366$ & 15 (3) & 16 (4) & 11 (4) & 7 (4) & 3 (3) & 4 (3) & 5 (3) & 7 (4)\\ 
			\hline
	\end{tabular}
\end{table}

Concerning the number of different models, results related to lasso regression are the same as those  concerning the number of different supports. Only the result for sd$\left(\epsilon\right)= 0.5$ differs, since one trial finished with a fifth value of the sparsity parameter and the same support as the model that was selected. Therefore, only results concerning models established using  SPLS methods are summarized in  Table \ref{tab.6}. In the case of bootstrap-based techniques, supports and models are similar, since no sparsity parameter is needed.

\begin{table}[ht]
	\caption{Results for SPLS model stability.  \label{tab.6}}
  \centering
		\begin{tabular}{lrrrr}
			\hline\hline
			 & \multicolumn{2}{c}{\# $\left(\eta_{opt},K_{opt}\right)$ $\left(\%\left(\eta,K\right)_{sel}\right)$} & \multicolumn{2}{c}{$\Gamma_2$ $\left(\%\mathcal{M}_{sel}\right)$}\\
       & \multicolumn{1}{c}{SPLS CV} & \multicolumn{1}{r}{SPLS BootYT} & \multicolumn{1}{c}{SPLS CV} & \multicolumn{1}{c}{SPLS BootYT}\\
		\hline
  sd$\left(\epsilon\right)= 0.5$ & 9 (46) & 3 (81) & 8 (55) & 2 (90) \\
	sd$\left(\epsilon\right)= 1$ & 13 (36) & 2 (73) & 9 (45) & 1 (100) \\ 
	sd$\left(\epsilon\right)= 3$ & 13 (17) & 5 (59) & 5 (34) & 4 (90) \\ 
	sd$\left(\epsilon\right)= 4$ & 15 (27) & 6 (37) & 12 (48) & 4 (67)\\ 
	sd$\left(\epsilon\right)= 5$ & 20 (19.33) & 6 (45) & 12 (25.33) & 4 (45)\\ 
	sd$\left(\epsilon\right)= 6.366$ & 16 (18) & 4 (46) & 11 (18) & 4 (46)\\ 
			\hline
	\end{tabular}
\end{table}

Concerning the three bootstrap-based techniques and in the light of accuracy results (Table \ref{tab.3}),  BootYT outperforms both the others, except in the case where sd$\left(\epsilon\right)= 0.5$, for which  BIC should be used. This exception is confirmed through the comparison of the dynamic bootstrap-based methods, where the sd$\left(\epsilon\right)= 0.5$ case is the only one where using the BIC criterion also represents the most relevant choice. This phenomenon matches with conclusions obtained by \citet{magnanensibootcrit}, in that the BIC criterion is well-designed for small values of noise variance, while  BootYT  outperforms it for non-negligible levels of random noise. The use of the $Q^2$ criterion for selecting the number of components would appear never be a reliable option, so  combining this criterion with our new dynamical approach was not done. The accuracy values highlight the fact that our new  method is always an improvement over the original one. Concerning the two versions of SPLS regression, both gave similar accuracy. These stand to perform better than others for the smallest levels of random noise variance, while  BootYTdyn outperforms all others for the highest values of noise variability. \\

Based on results introduced in Tables \ref{tab.4} and \ref{tab.6}, the $Q^2$-based approach for predictor selection has non-negligable stability issues, providing between 18 and 96 different models in the 100 simulations. Depending on both the noise variability and the criterion combined with our dynamic approach, the latter improves over the original one by stabilizing the choice of sparse support. Cases where this is observed match with previous conclusions established when analyzing accuracy results, namely both the BIC-based dynamic approach for small values of noise variability, and  BootYTdyn used for datasets with non-negligible noise variances. This strengthens  the conclusion that  BIC  is well-designed for small values of noise variance, while  BootYT  performs best for non-negligible noise. As for SPLS methods, our new bootstrap-based version gains in stability in that it retains fewer different optimal pairs $\left(\eta_{opt},K_{opt}\right)$, and also sparse models, than the original does. Moreover, since $\Gamma_1=\Gamma_2$ for all studied datasets, this directly implies that it retains a unique optimal number of components for each sparse support. It thus permits a choice of optimal model in a more reliable way than using the CV-based technique. Lastly, lasso regression has  good stability performances, leading it,  BootYTdyn and  SPLS BootYT to be recommended when stability is important.\\

\begin{figure}[ht]
		\centering
          \includegraphics[trim = 0cm 0cm 0cm 0cm, clip,scale=0.7]{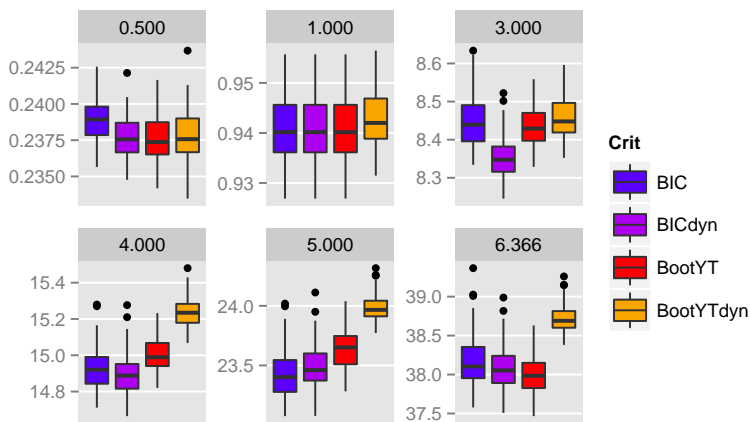}
    \caption{Boxplots of 10-CV MSE based on $\mathbf{y}$ with noise.}
   \label{fig.5}
\end{figure}

The last descriptive statistic concerns the number of significant predictors retained (Table \ref{tab.5}). The $Q^2$-based approach is the least sparse, selecting the highest number of covariates. Both  BICdyn  BootYTdyn  improve respectively  BIC and BootYT, showing  better accuracy related to their selected support. Indeed, the four expected covariates are always included in these selected supports. Once more,  BootYT  can be recommended for datasets with non-negligible random noise variability compared with corrected BIC, which has to be applied for small values of random variance. Indeed, both the BIC and BICdyn approaches, by retaining globally increasing numbers of predictors while the random noise standard deviation increases, lead us to suppose that they use some predictors to model this random noise, leading to over-fitting. On the contrary, while  BootYT  selects a stable number of significant predictors,  BootYTdyn selects a decreasing number of significant predictors as the random noise standard deviation increases, which is expected. As a confirmation of over-fitting issues with BIC, we present the 10-fold CV-based MSE in Fig. \ref{fig.5}.

These results tend to confirm our suspicion of over-fitting since, except for results related to BootYTdyn, the others have MSE that do not match with the theoretical random noise variances, suggesting that  a part of the noise is being modeled. As for the two SPLS methods, there is no pertinent difference to mention, both of them concluding on similar numbers of selected predictors. Lastly, like in BIC-based approaches, the lasso  extracts an increasing number of significant predictors.\\

As a first conclusion, we can thus reasonably conclude that, based on these first simulation results,  BootYTdyn and SPLS BootYT should be used in practice.

\subsubsection{Complexity and predictive ability results}
\label{325}

To confirm and strengthen the conclusions of Section \ref{322}, we will now focus on the predictive abilities of the models selected by the various approaches. We calculate one hundred times the 10-fold CV MSE based on the original simulated response values (without noise) of the various selected models. These results thus reflect the accuracy in predicting the original information by leaving out random noise. We also compute the Predictive MSE (PMSE) based on the test set, by using its simulated response $\mathbf{y_\text{test}}$ without including noise. Lastly, we extract the DoF of each selected sparse model ($\text{DoF}_{sel}$) to compare the respective complexities.\\ 

\begin{figure}[h]
		\centering
        \subfigure{\includegraphics[trim = 0cm 0.2cm 3.3cm 0cm, clip,scale=0.37]{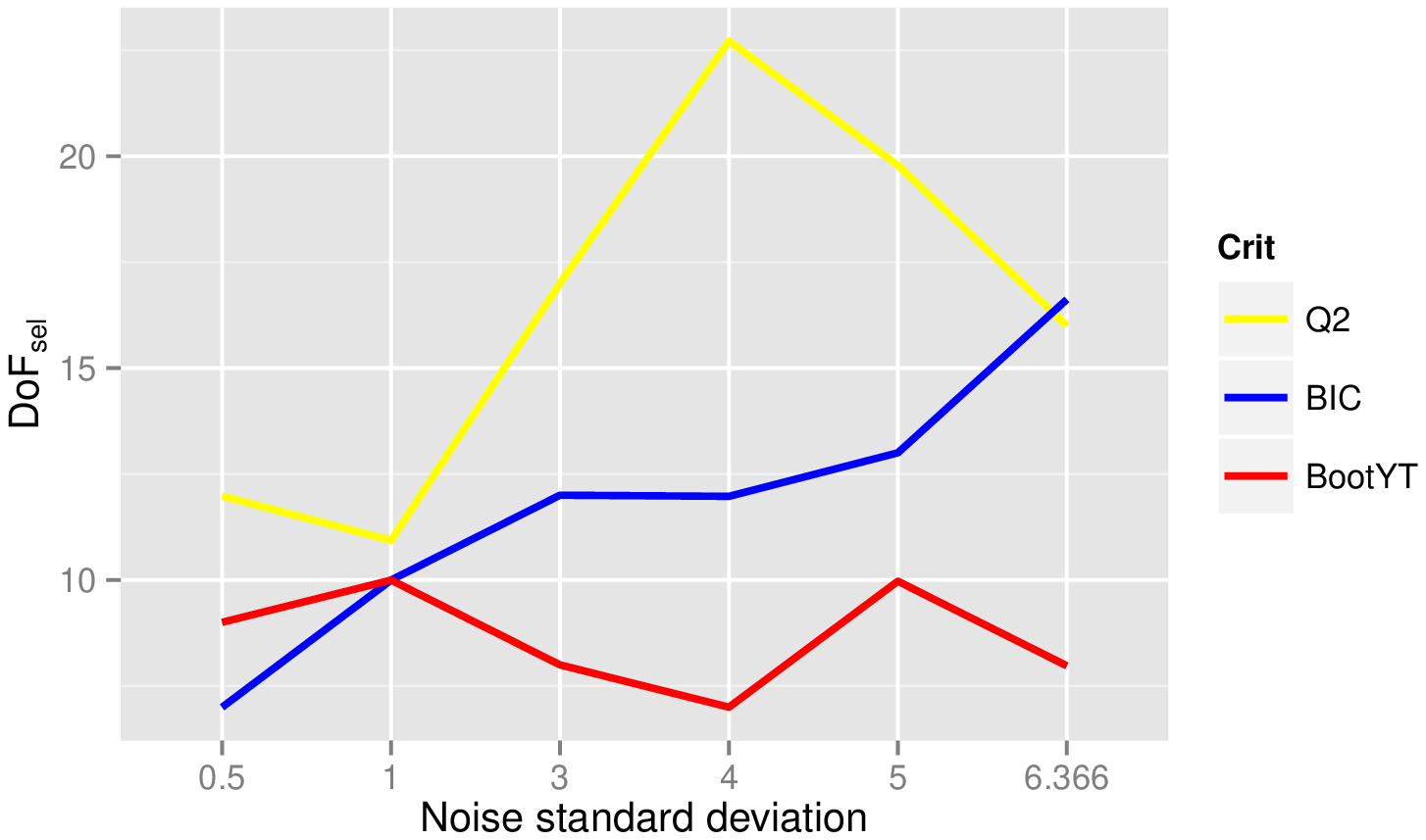}}
        \subfigure{\includegraphics[trim = 0cm 0.2cm 1cm 0cm, clip,scale=0.37]{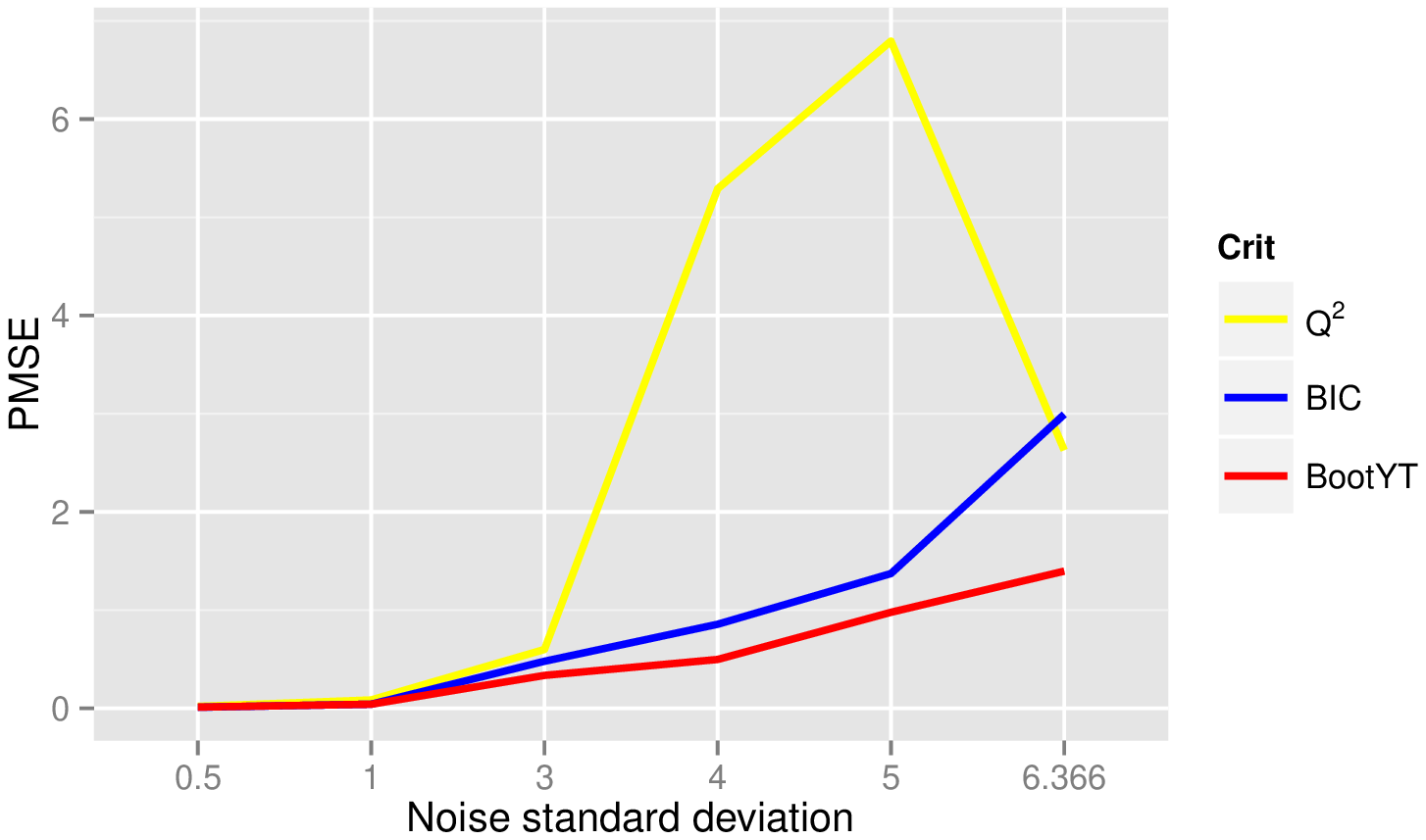}}
        \subfigure{\includegraphics[trim = 0cm 0.2cm 1cm 0cm, clip,scale=0.37]{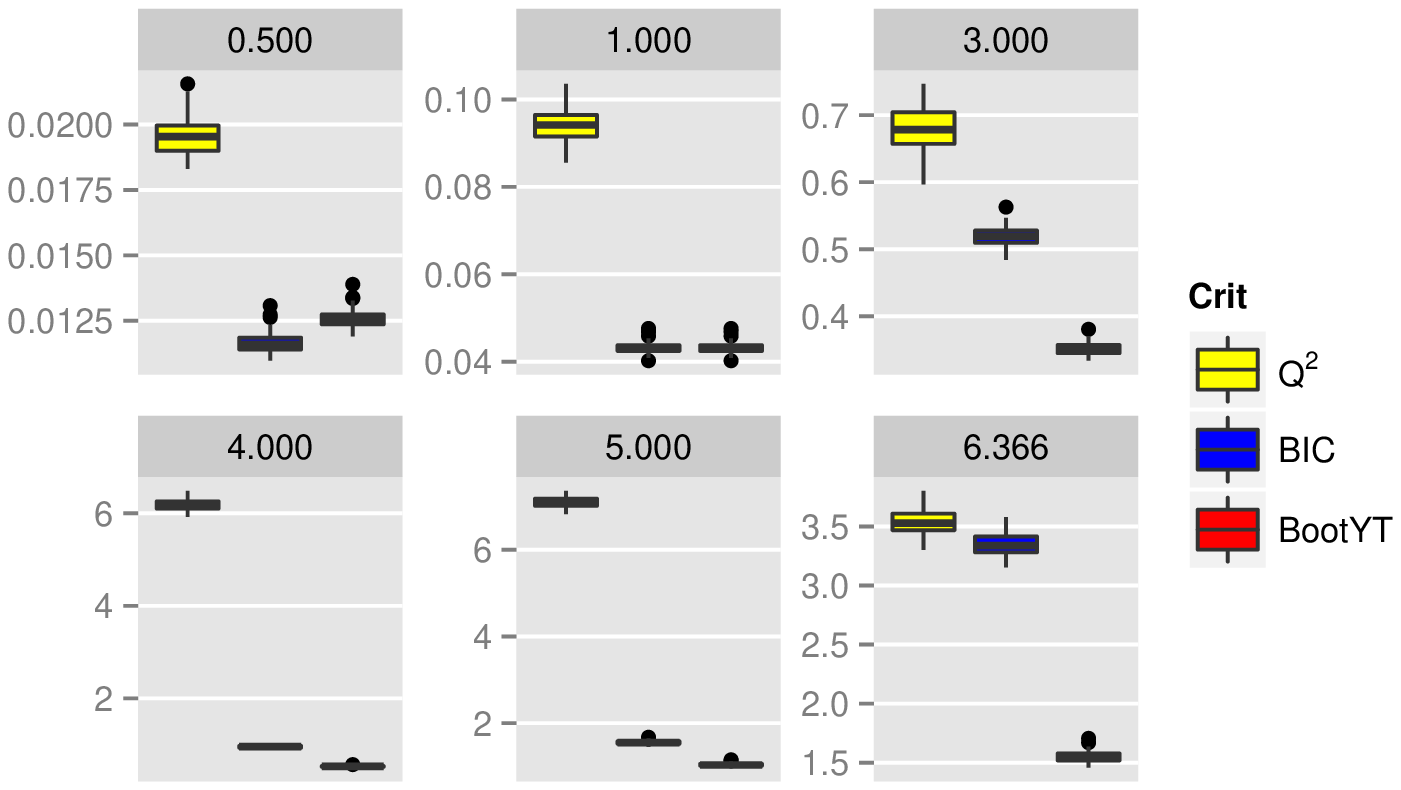}}
				\caption{From left to right: DoF of the extracted sparse models, PMSE based on $\mathbf{y}_\text{test}$ without noise, and boxplots of 10-CV MSE based on $\mathbf{y}$ without noise.}
   \label{fig.4}
\end{figure}

Graphical results for these statistics, related to $Q^2$, BIC and BootYT,  are shown in Fig.~\ref{fig.4}.  

The evolution of estimated DoF both highlights and confirms  BIC and $Q^2$ over-fitting issues. Indeed, as the random noise standard deviation increases, these methods globally build sparse models with increasing complexity. Thus, they model  increasing parts of the inserted random noise, implying poor predictive abilities of their selected models compared to those obtained by applying BootYT. This is confirmed through higher values of PMSE and CV MSE, especially for datasets with non-negligible random noise variability. These results confirm the conclusions from the previous section in that using the $Q^2$ criterion for selecting the number of PLS components should be avoided, and that BootYT  outperforms corrected BIC, except for responses with negligible random noise levels. Therefore, only  BIC and BootYT are retained for further comparison.

\begin{figure}[h]
		\centering
		    \subfigure{\includegraphics[trim = 0cm 0.2cm 3.9cm 0cm, clip,scale=0.37]{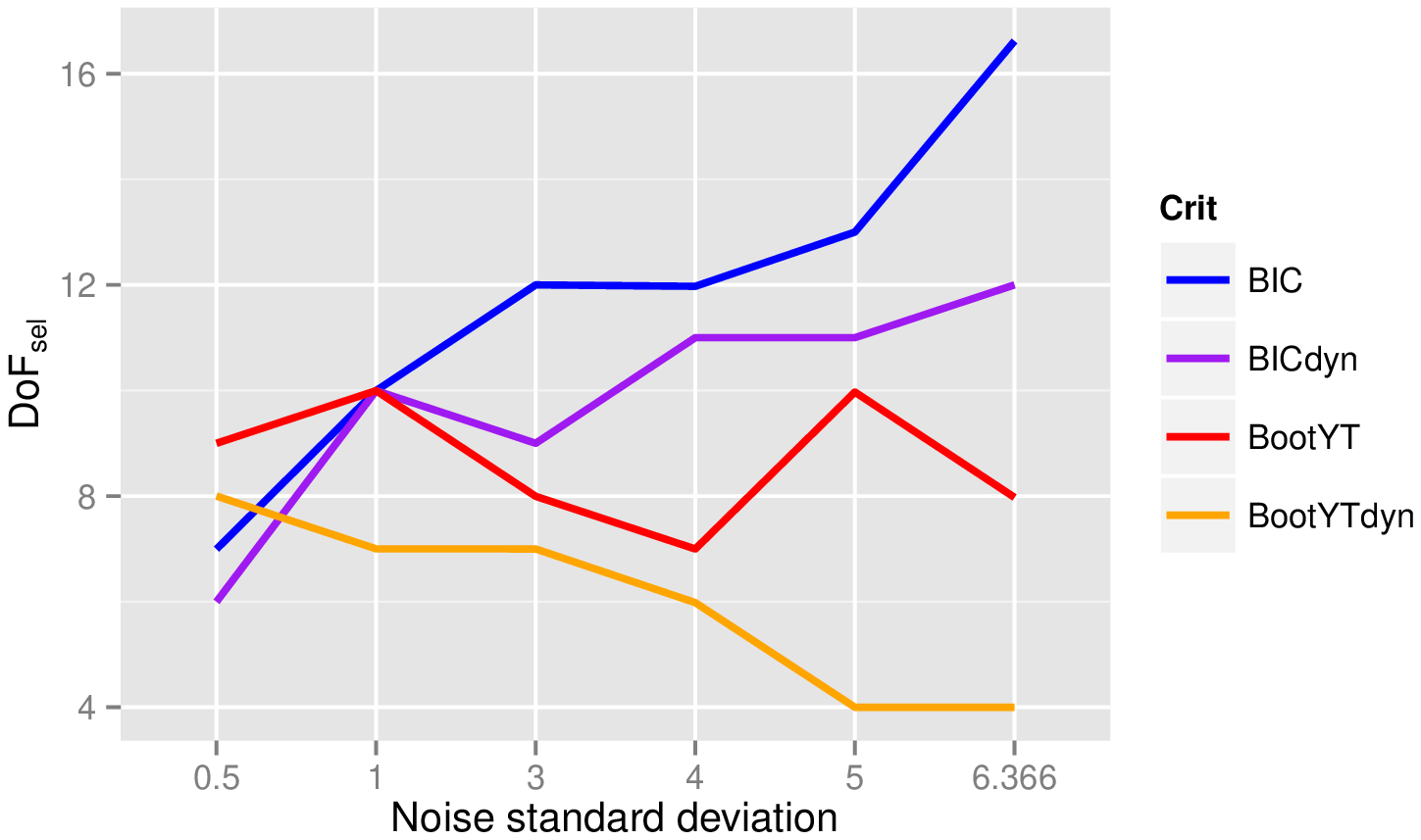}}
        \subfigure{\includegraphics[trim = 0cm 0.2cm 1cm 0cm, clip,scale=0.37]{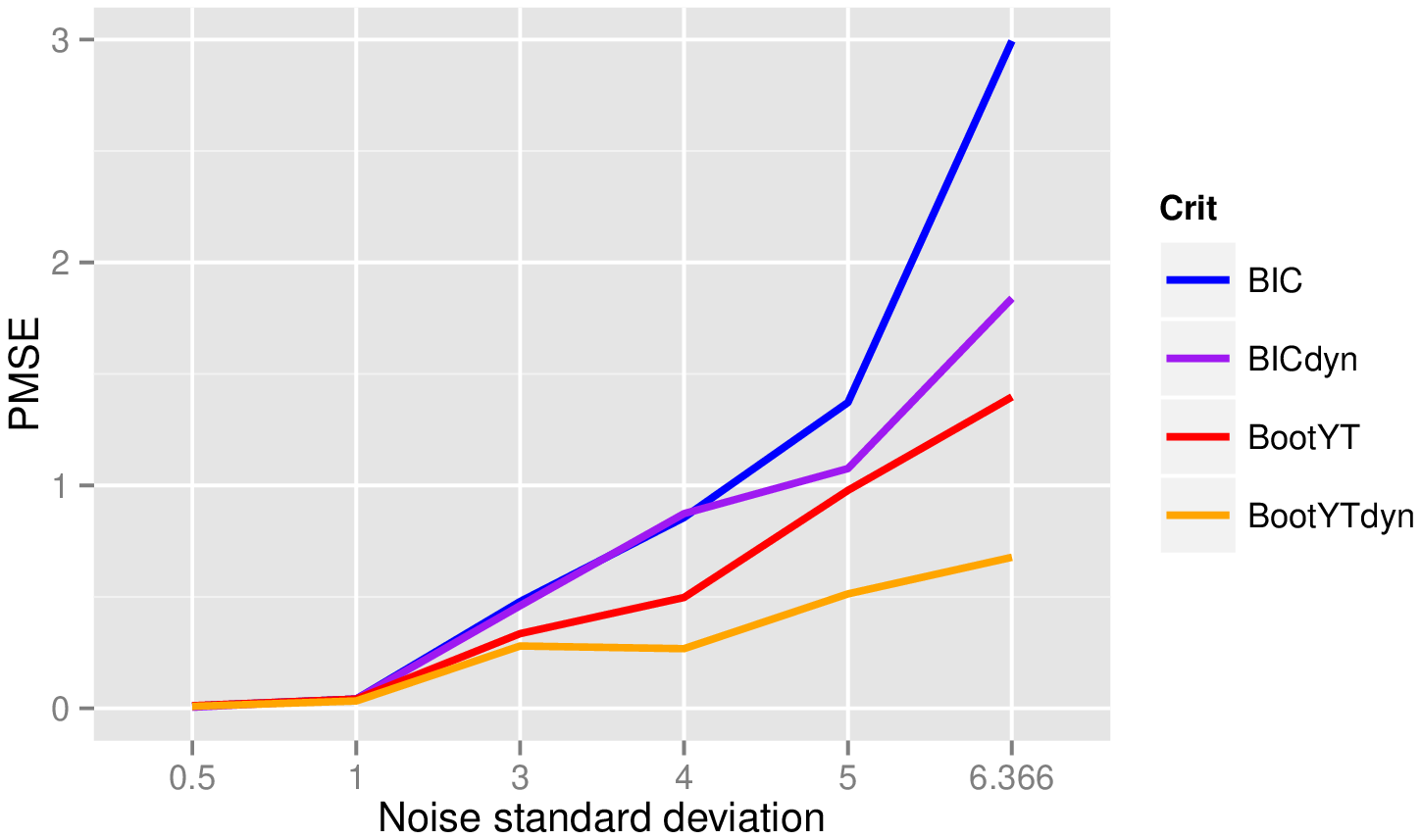}}
        \subfigure{\includegraphics[trim = 0cm 0.2cm 1cm 0cm, clip,scale=0.37]{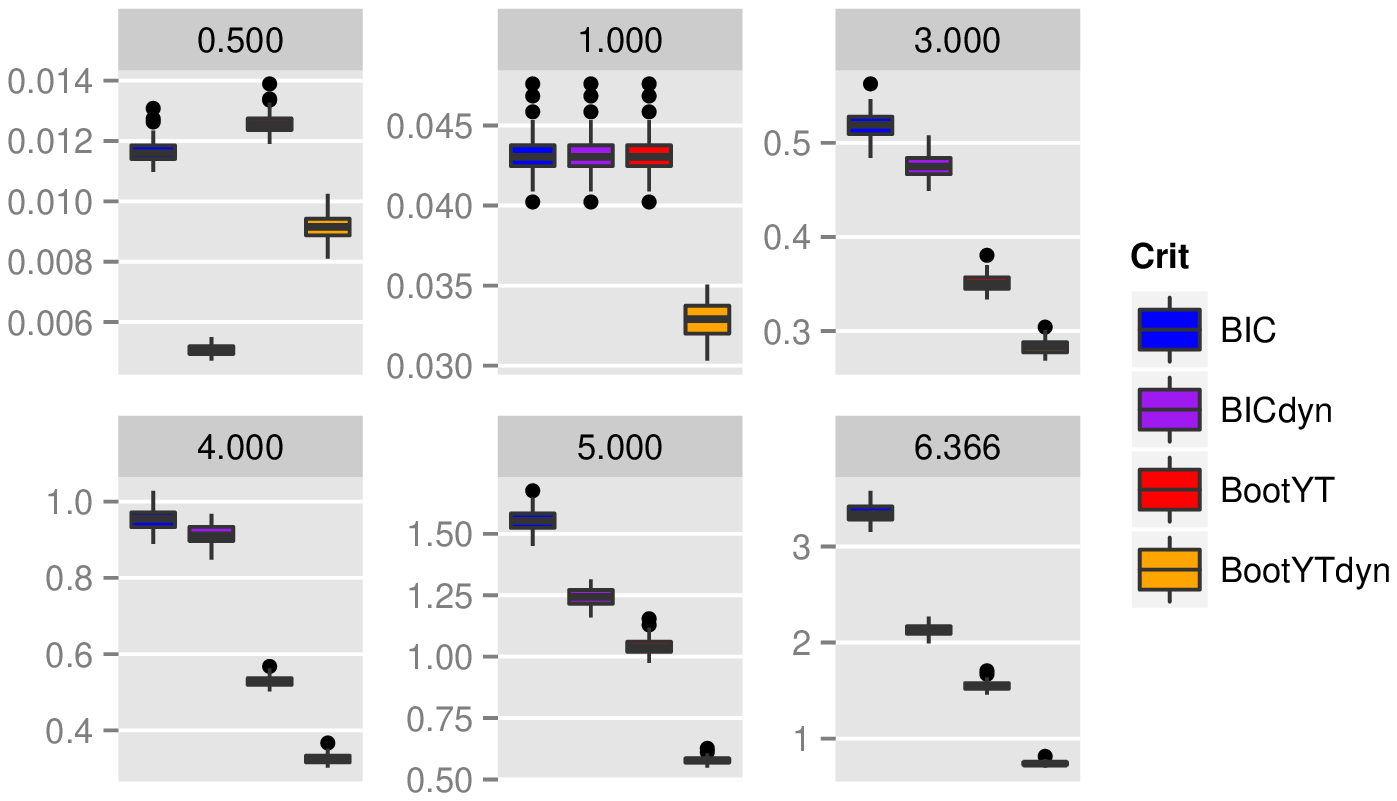}}\\
				\caption{From left to right: DoF of the extracted sparse models, PMSE based on $\mathbf{y}_\text{test}$ without noise, and boxplots of 10-CV MSE based on $\mathbf{y}$ without noise.}
   \label{fig.6}
\end{figure}

The results shown in Fig.~\ref{fig.6} highlight that BootYTdyn  is the only method that models with decreasing DoF, ensuring a complexity reduction suitable to the avoidance of prediction issues. Indeed, BootYTdyn selects models with the lowest PMSE and 10-CV MSE.

\begin{figure}[h]
		\centering
				\subfigure{\includegraphics[trim = 0cm 0.2cm 4.3cm 0cm, clip,scale=0.38]{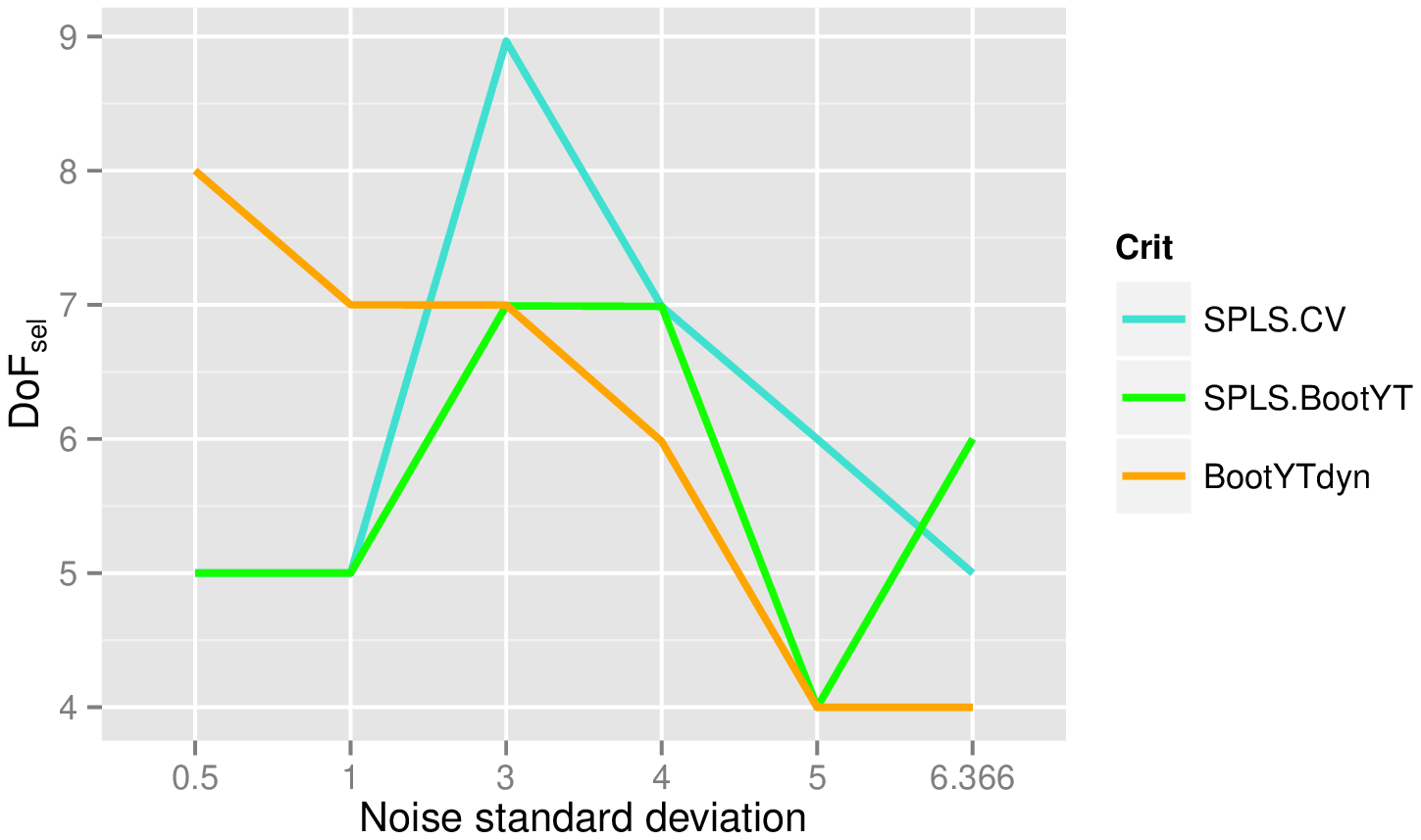}}
        \subfigure{\includegraphics[trim = 0cm 0.2cm 1cm 0cm, clip,scale=0.38]{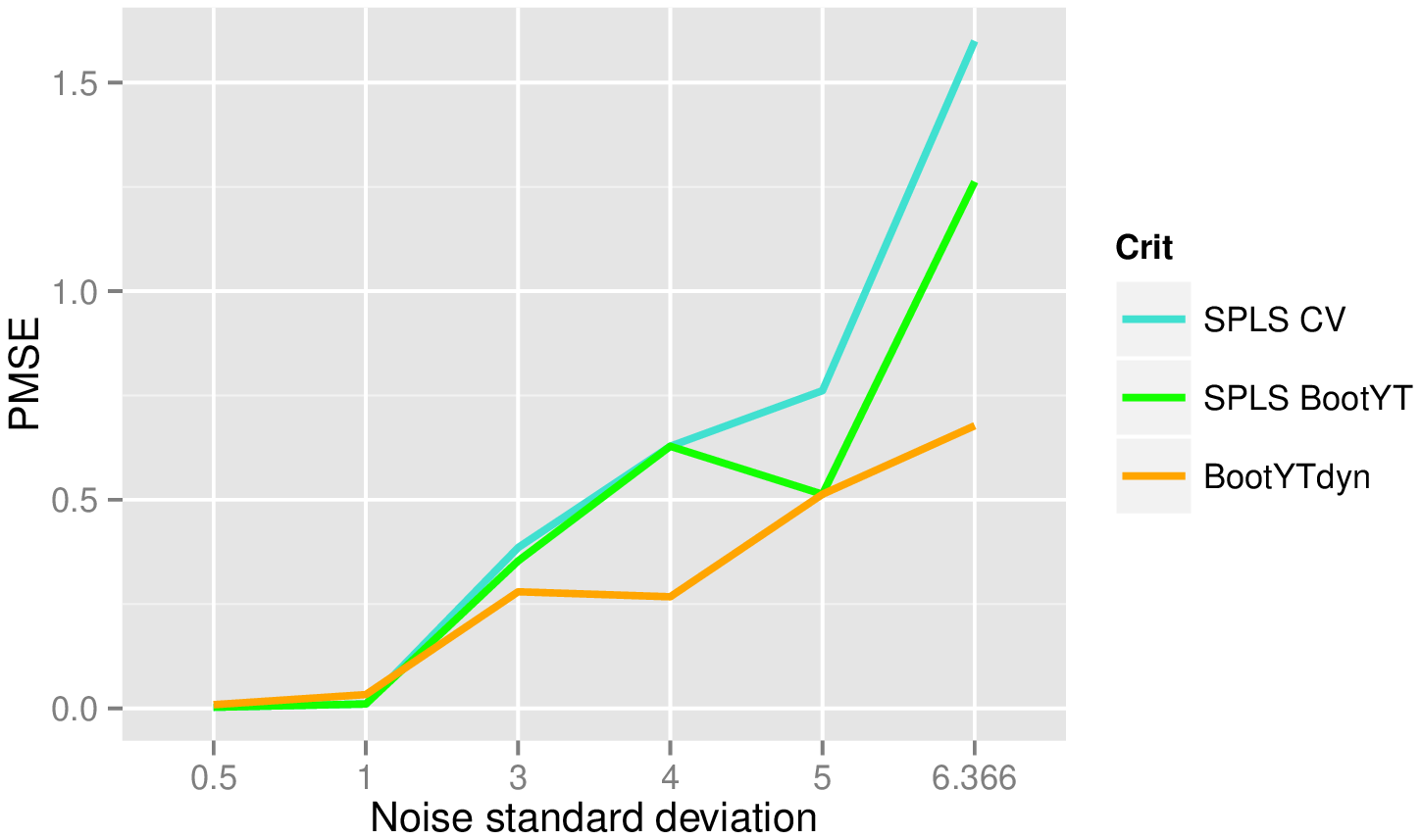}}
        \subfigure{\includegraphics[trim = 0cm 0.2cm 1cm 0cm, clip,scale=0.38]{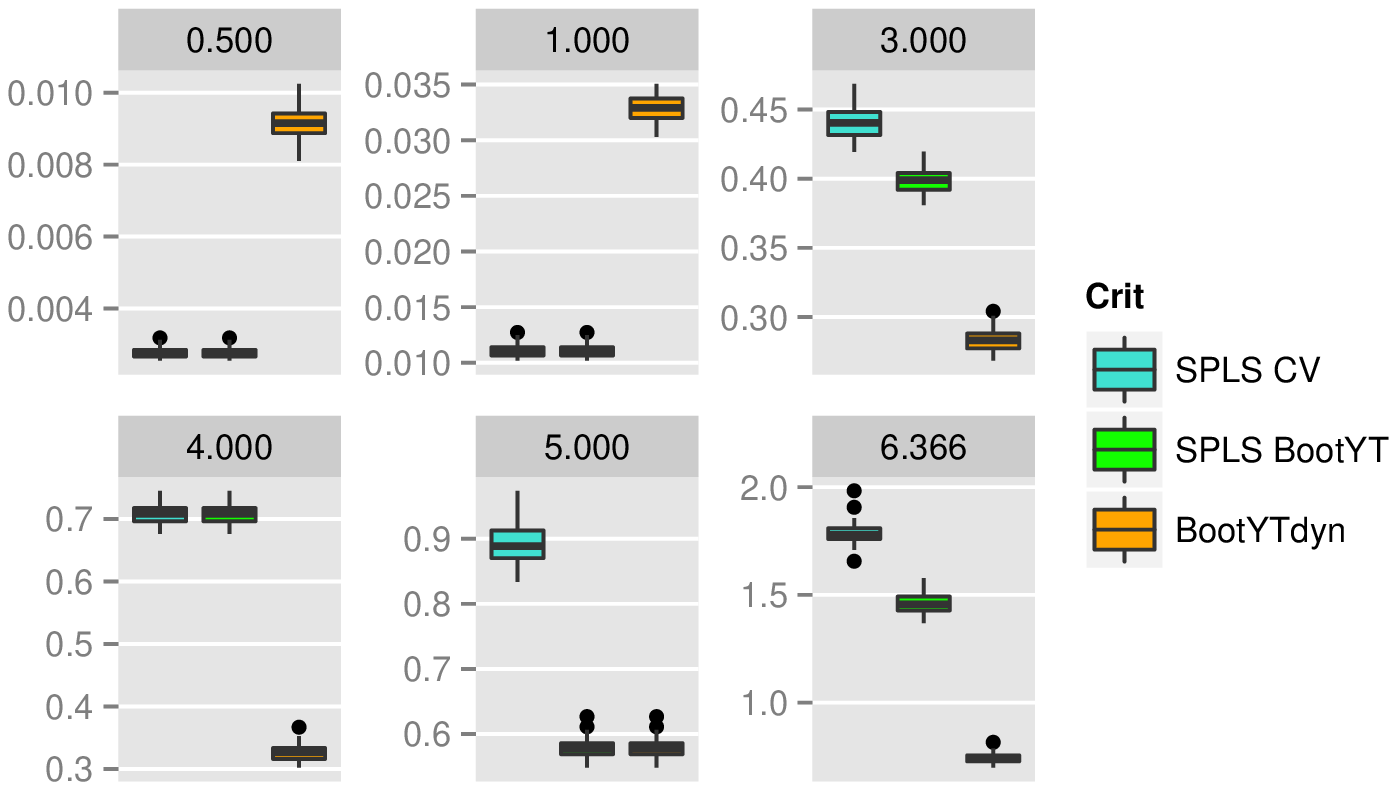}}\\
				\caption{From left to right: DoF of the extracted sparse models, PMSE based on $\mathbf{y}_\text{test}$ without noise, and boxplots of 10-CV MSE based on $\mathbf{y}$ without noise.}
   \label{fig.7}
\end{figure}

In light of these results, only  BootYTyn  is retained for further comparison. Comparing the two SPLS implementations with respect to their predictive abilities lead us to recommend SPLS BootYT, since models selected by this bootstrap-adapted SPLS technique feature comparable if not lower PMSE and 10-CV MSE (Fig.~\ref{fig.7}). Let us clarify that, in order to ensure relevant comparisons, we used ordinary PLS regressions with both the support and the number of components selected by the SPLS methods, and not SPLS methods with selected tuning parameters, for computing the 10-CV MSE.

For datasets characterized by a low level of random noise variability in $\mathbf{y}$,  SPLS BootYT builds models inducing the smallest CV MSE and PMSE. By focusing on non-negligible random variability,  BootYTdyn gives the smallest predictive errors,  thus showing the high level of robustness of this approach against  random noise. This robustness is, as previously mentioned, due to its decreasing number of significant predictors, and also components, leading to decreasing DoF, i.e., a loss of complexity.

\begin{figure}[h]
		\centering
				\subfigure{\includegraphics[trim = 0cm 0.2cm 4.3cm 0cm, clip,scale=0.38]{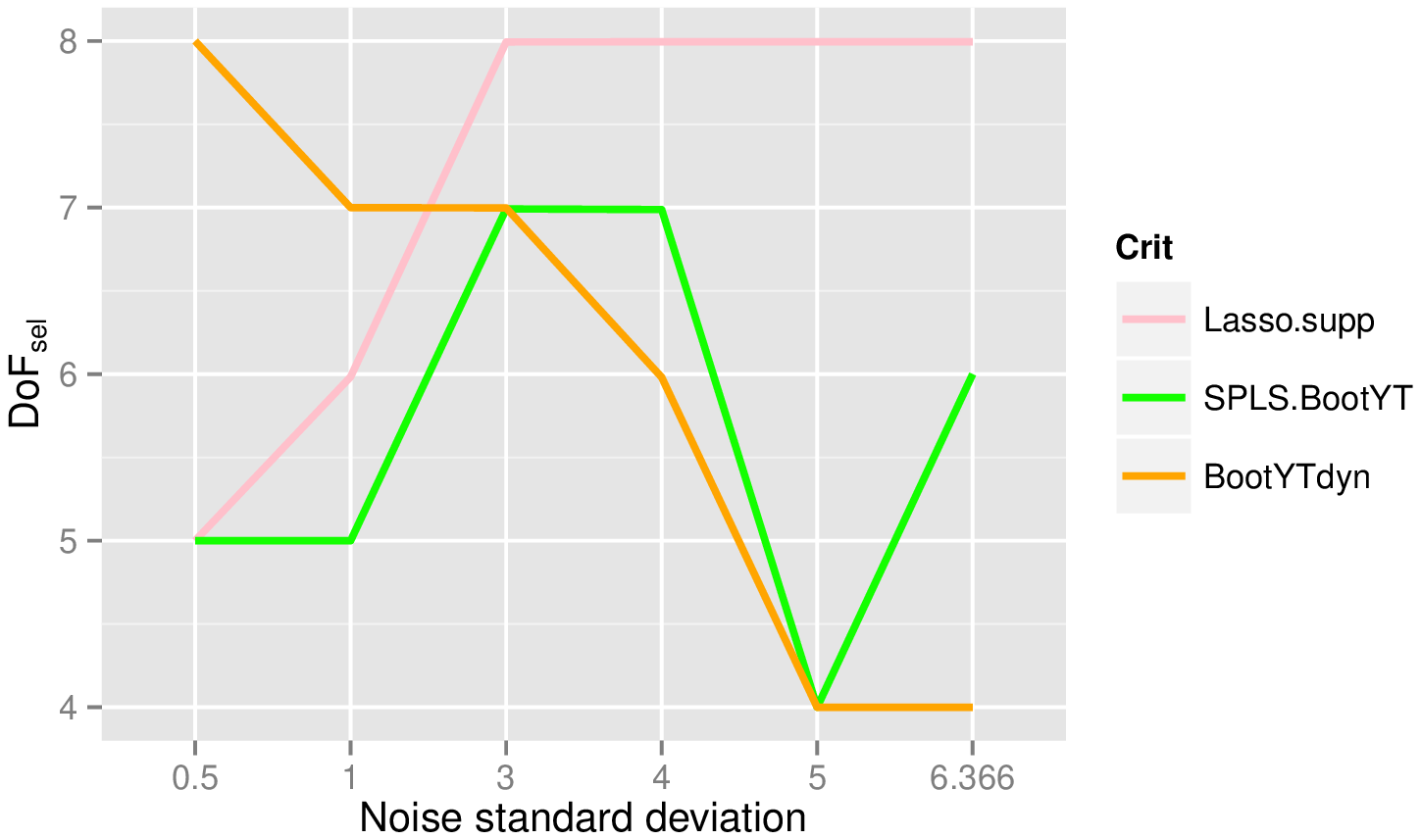}}
        \subfigure{\includegraphics[trim = 0cm 0.2cm 1cm 0cm, clip,scale=0.38]{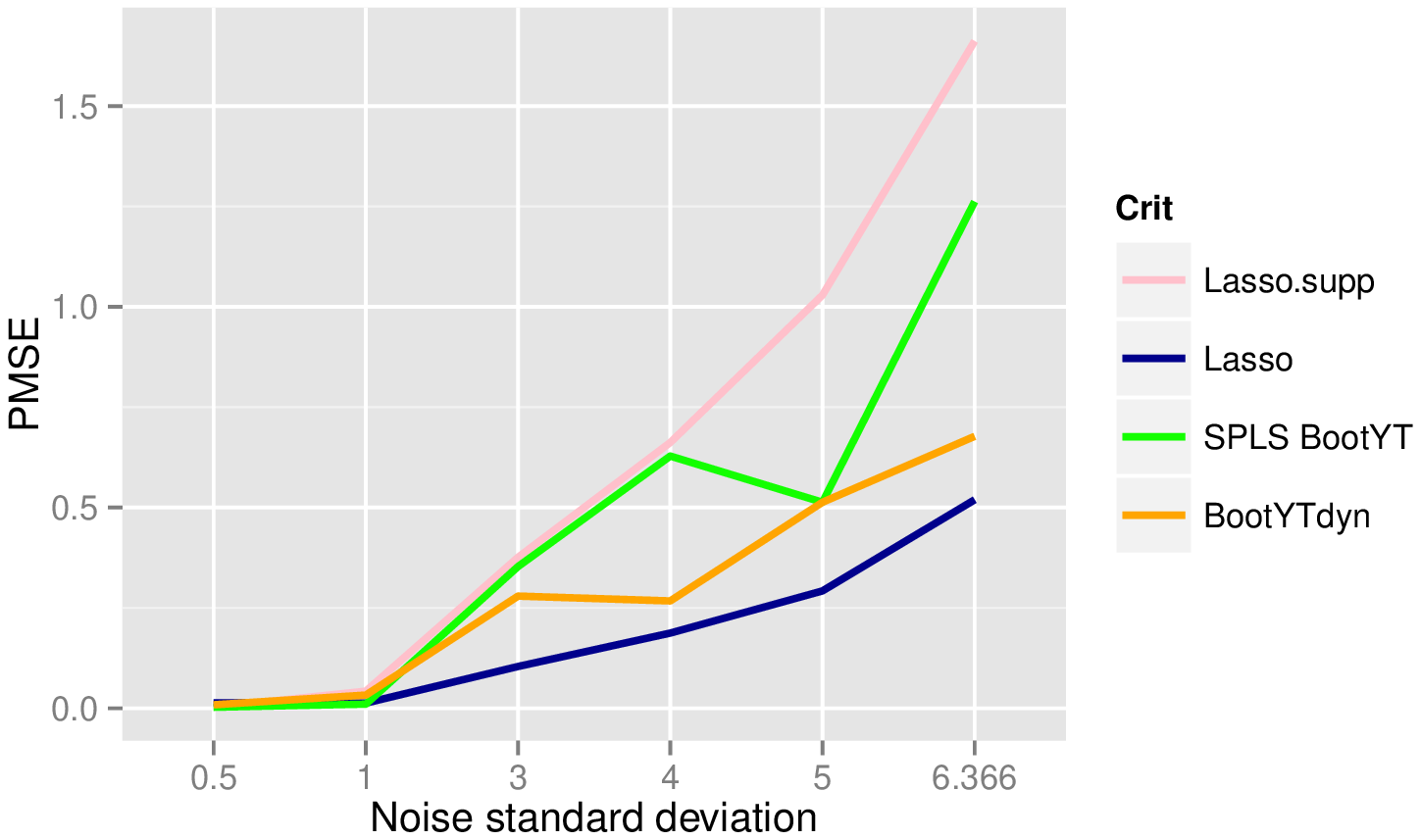}}
        \subfigure{\includegraphics[trim = 0cm 0.2cm 1cm 0cm, clip,scale=0.38]{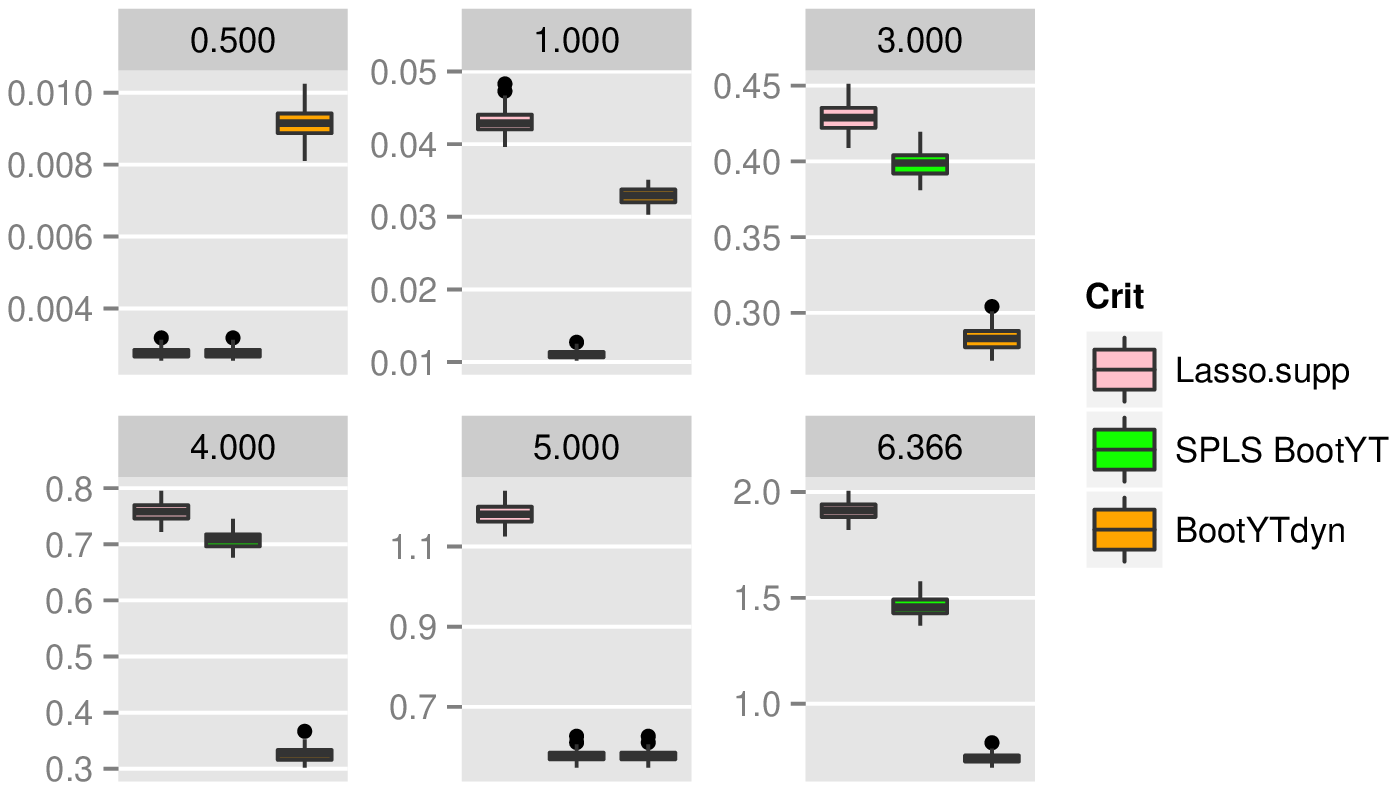}}\\
				\caption{From left to right: DoF of the extracted sparse models, PMSE based on $\mathbf{y}_\text{test}$ without noise, and boxplots of 10-CV MSE based on $\mathbf{y}$ without noise.}
   \label{fig.8}
\end{figure}

Lastly, we compare the two retained methods BootYTdyn and SPLS BootYT, and the lasso is run. As for SPLS methods, in order to perform relevant comparisons, the supports extracted by the lasso  are used as sets of covariates for a PLS regression. The number of PLS components is then established by performing one hundred times their selection using the bootstrap-based stopping criterion; the number of components related to the highest occurrence rate was selected. In this way, results shown in Fig.~\ref{fig.8} concerning  10-CV MSE give the predictive ability of the extracted supports for PLS regression. In order to provide a clear picture of the impact of this choice, the PMSE obtained through the lasso  is also shown in Fig.~\ref{fig.8}. These approaches are referred to as Lasso and Lasso.supp.

Except for negligible values of random noise variability in $\mathbf{y}$, the support extracted by the lasso regression and applied as explanatory variables for PLS regression are linked to both the highest 10-CV MSE and  highest PMSE. This is a direct consequence of the lasso's accuracy issues mentioned in Section \ref{324}, notably the increasing number of extracted covariate while the random noise variability rises. However, performing  predictions using the model obtained by the original lasso regression lead to the lowest PMSE values. This is due to the $L^1$ penalization which is applied on the vector of estimated parameters, thus permitting correction of the relative lack of accuracy of this technique.\\ 

To conclude, we summarize our conclusions in the following Table \ref{tab.8}, and recommend certain approaches, depending on whether the initial aim was to select significant predictors or to obtain a sparse model with attractive predictive ability.

\begin{table}[h]
	\caption{Recommended approaches.  \label{tab.8}}
  \centering
		\begin{tabular}{lcc}
			\hline\hline
       & \multicolumn{1}{c}{Accuracy} & \multicolumn{1}{c}{Predictive ability} \\
		\hline
  Low noise variability & SPLS BootYT & SPLS BootYT \\
	High noise variability & BootYTdyn & Lasso$/$BootYTdyn \\ 
			\hline
	\end{tabular}
\end{table}

\section{Real dataset application}
\label{4}

In this section, we deal with the predictors matrix introduced in Section \ref{321} and the original binary response vector.  Five approaches for variable selection, adapted for the GPLS framework, are considered for comparison.
\begin{enumerate}
\item \textbf{BootYT}. The bootstrap-based method, introduced by \citet{lazraq2003selecting}, combined with the bootstrap-based criterion \citep{magnanensibootcrit} for pre-selecting the number of components.
\item \textbf{BootYTdyn}. The new dynamic bootstrap-based method combined with the bootstrap-based criterion for successive determinations of the number of components.
\item \textbf{SGPLS CV}. The original SGPLS method using 10-fold CV for tuning parameter selection \citep{chung2010sparse}.
\item \textbf{SGPLS BootYT}. The new adapted SGPLS version using the bootstrap-based criterion for component selection.
\item \textbf{RSGPLS}. An approach developed by \citet{durif2015adaptive}. It consists of adapting the SGPLS method by introducing a ridge penalty to ensure  convergence of parameter estimations, and stability in hyper-parameter tuning. They also propose an adjustment of the $L_1$ constraint in order to further penalize  less significant predictors. Hyper-parameters are tuned using CV MSE.
\item \textbf{Lasso}. The adapted lasso regression for logistic framework, available in the R package \textit{glmnet}, as a benchmark.
\end{enumerate}

Concerning bootstrap-based approaches, the incorporation of  PLS methodology into GLM, developed by \citet{bastien2005pls}, was used. Due to non-convergence issues for parameter estimations, some samples were excluded using a threshold for parameter estimations. Indeed, a model built using a bootstrap sample with at least one parameter estimate that is higher in absolute value than $10^4$ times the one (in absolute value as well) estimated on the original dataset, leads to the exclusion of this bootstrap sample. Thus, to ensure a sufficient number of relevant bootstrap samples, the preset number of computed samples was increased to $R=4000$. As for the lasso, three different loss functions for the establishment of the sparsity parameter using 10-fold CV were used: the number of misclassified values, the MSE, and the deviance. We  refer to them as Lasso.Cl, Lasso.MSE and Lasso.Dev, respectively. The set of  values defined for the sparsity parameter is preset by the ``glmnet'' package. For the SGPLS and RSGPLS approaches, the number of components $K$ varies from 1 to 10, the sparsity parameter $\eta$ varies from 0.04 to 0.99 (by steps of 0.05), and the ridge parameter involved in the RSGPLS technique is selected from 31 points  $\text{log}_{10}$-linearly spaced in the range $\left[10^{-2},10^3\right]$, as in \citet{durif2015adaptive}. 

Each method was performed one hundred times in order to obtain relevant results. However, due to the high observed stability of results extracted with the BootYTdyn approach, and in order to save computational time, this was only performed twenty times instead of a hundred.\\ 

\begin{figure}[ht]
	\centering
	\includegraphics[trim = 0cm 0cm 0cm 0cm, clip,scale=0.7]{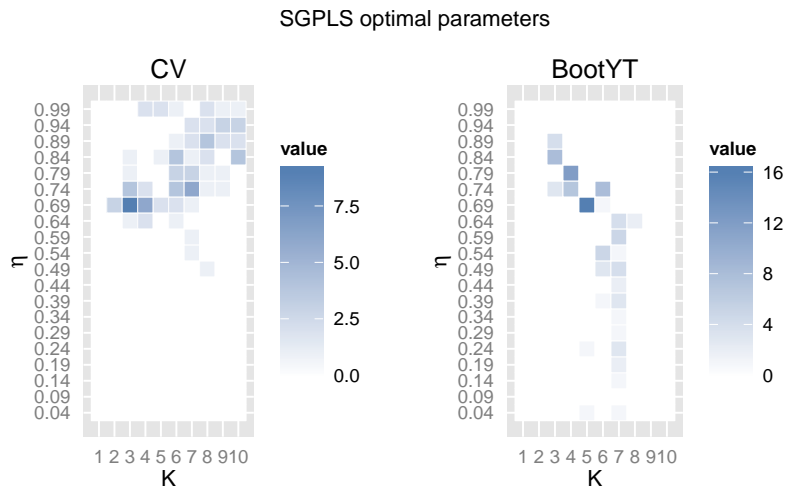}
	\caption{Repartition of selected sets of tuning parameters over the 100 trials, for SGPLS CV (left) and  SGPLS BootYT (right).}
	\label{fig.6.9}
\end{figure}

Stability results for tuning parameter selection, except for the bootstrap-based methods, are shown in Fig.~\ref{fig.6.9} and Table~\ref{tab.6.9}. 

\begin{table}[ht]
	\caption{Number of different selected sets of tuning parameters over the 100 trials (rate of occurrence of the retained set of tuning parameters).  \label{tab.6.9}}
	\centering	
		\begin{tabular}{rrrrrrrr}
		\hline\hline
			\multicolumn{1}{c}{SGPLS CV} &\multicolumn{1}{c}{SGPLS BootYT} & \multicolumn{1}{c}{RSGPLS} & \multicolumn{1}{c}{Lasso.Cl} & \multicolumn{1}{c}{Lasso.MSE} & \multicolumn{1}{c}{Lasso.Dev}\\
			\hline
			44 (9) & 26 (16) & 64 (5) & 33 (8) & 6 (44) & 5 (48)\\
			\hline
		\end{tabular}
\end{table}

Based on results summarized in Table \ref{tab.6.9}, the lasso methodology, using CV-based MSE or deviance values for the selection of its hyper-parameter, is the most stable. This can be explained by the fact that only one  hyper-parameter is involved, while the other techniques are based on two or three tuning parameters. Using the number of misclassified values for CV has to be avoided for stable selection of the sparsity parameter. Our SGPLS adaptation with the BootYT criterion improves  reliability in selecting the set of tuning parameters, as previously observed in the PLS framework (Sections \ref{22} and \ref{324}). Note that, concerning  RSGPLS, three different sets of optimal parameters were extracted with maximal occurrence rate of five, all of them selecting the same set of predictors. Thus, the set of parameters which leads to the smallest number of misclassified values on the training dataset was retained. As already mentioned in Section \ref{323}, extracting the same support does not necessarily lead to the same model when sparsity or ridge parameters are involved. Therefore, the numbers of sparse supports and models retained are respectively summarized in Tables~\ref{tab.6.10} and \ref{tab.6.11}.

\begin{table}[h]
	\caption{Number $\Gamma_1$ of different extracted supports $\left(\%\mathcal{S}_{sel}\right)$.  \label{tab.6.10}}
	\centering	
		\begin{tabular}{rrrrrrrr}
		  \hline\hline
			\multicolumn{1}{c}{SGPLS CV} &\multicolumn{1}{c}{SGPLS BootYT} & \multicolumn{1}{c}{RSGPLS} & \multicolumn{1}{c}{Lasso.Cl} & \multicolumn{1}{c}{Lasso.MSE} & \multicolumn{1}{c}{Lasso.Dev} & \multicolumn{1}{c}{BootYT} & \multicolumn{1}{c}{BootYTdyn}\\
			\hline
			44 (9) & 14 (16) & 26 (40) & 26 (13) & 5 (83) & 4 (83) & 4 (35) & 2 (80)\\
			\hline
		\end{tabular}
\end{table}

\begin{table}[h]
	\caption{Number $\Gamma_2$ of different extracted sparse models $\left(\%\mathcal{M}_{sel}\right)$.  \label{tab.6.11}}
	\centering	
		\begin{tabular}{rrrrrrrr}
		\hline\hline
			\multicolumn{1}{c}{SGPLS CV} &\multicolumn{1}{c}{SGPLS BootYT} & \multicolumn{1}{c}{RSGPLS} & \multicolumn{1}{c}{Lasso.Cl} & \multicolumn{1}{c}{Lasso.MSE} & \multicolumn{1}{c}{Lasso.Dev} & \multicolumn{1}{c}{BootYT} & \multicolumn{1}{c}{BootYTdyn}\\
			\hline
			44 (9) & 16 (16) & 60 (5) & 33 (8) & 6 (44) & 5 (48) & 4 (35) & 2 (80)\\
			\hline
		\end{tabular}
\end{table}

In  light of these results,  BootYTdyn and both the MSE- and deviance-based lasso techniques are the most stable in extracting supports and models. This could be due in part to the fact than all of them depend on only one tuning parameter. Even if this hyper-parameter for the BootYTdyn approach, i.e., the number of components, has to be chosen $R$ times, the high stability of the bootstrap-based stopping criterion introduced by \citet{magnanensibootcrit} endows this approach with good  stability in selecting the sparse support. As for the PLS framework, our new bootstrap-based SGPLS implementation improves the stability here. The lack of stability of the lasso based on misclassified values is directly induced by the discrete form of this statistic. This issue was also observed and mentioned in \citet{magnanensibootcrit}. 

\begin{figure}[ht]
	\centering
	\includegraphics[trim = 0cm 3cm 0cm 3cm, clip,scale=1]{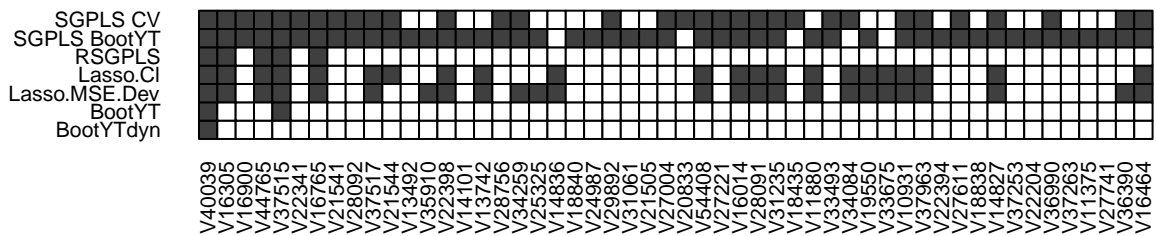}
	\caption{Summary of selected supports.}
	\label{fig.6.10}
\end{figure}

\begin{sloppypar}
The various selected supports are displayed in Fig.~\ref{fig.6.10}. Note that both the MSE- and deviance-based lasso regressions select the same support and the same model, noted  Lasso.MSE.Dev in the following.
\end{sloppypar}

In the following, two additional independent public datasets, stated by \citet{marisa2013gene} as being comparable with our original dataset, were included for comparison of predictive abilities. These datasets are named GSE18088 (n=53) \citep{grone2011molecular} and GSE14333 (n=247) \citep{jorissen2009metastasis}. Both the MSE and number of misclassified (MC) value of the selected models were computed on both the training and test parts of the original dataset, as well as on the two additional datasets. These results are summarized in Table~\ref{tab.6.12}. Since independent datasets were available, we decided to follow the suggestion of \citet[p.1596]{van2009survival}: ``the true evaluation of a predictor's performance is to be done on independent data''. 

\begin{table}[h]
	\caption{Summary of model fitting and predictive abilities.  \label{tab.6.12}}
	\centering	
		\begin{tabular}{lrrrrrrrr}
		 \hline\hline
			&\multicolumn{4}{c}{MC}&\multicolumn{4}{c}{MSE}\\
			\multicolumn{1}{l}{GSE} &\multicolumn{1}{c}{39582trai} &\multicolumn{1}{c}{39582test} & \multicolumn{1}{c}{18088} & \multicolumn{1}{c}{14333} & \multicolumn{1}{c}{39582trai} &\multicolumn{1}{c}{39582test} & \multicolumn{1}{c}{18088} & \multicolumn{1}{c}{14333} \\
			\hline
			SGPLS CV & 48 & 19 & 10 & 45 & 0.1372 & 0.1374 & 0.1743 & 0.1520\\
			SGPLS BootYT & 40 & 15 & 12 & 46 & 0.1295 & 0.1328 & 0.1848 & 0.1601\\
			RSGPLS & 53 & 13 & 17 & 45 & 0.1005 & 0.0912 & 0.1909 & 0.1276 \\
			Lasso.Cl & 50 & 11 & 12 & 44 & 0.0928 & 0.0827 & 0.1283 & 0.1425\\
			Lasso.MSE.Dev & 46 & 15 & 12 & 47 & 0.0875 & 0.0828 & 0.1242 & 0.1406\\
			BootYT  & 73 & 23 & 13 & 111 & 0.1272 & 0.1256 & 0.1817 & 0.2924 \\
			BootYTdyn & 92 & 25 & 13 & 35 & 0.1507 & 0.1446 & 0.1695 & 0.1325 \\
			\hline
		\end{tabular}
\end{table}

\begin{figure}[h]
	\centering
	\includegraphics[trim = 0cm 0cm 0cm 0cm, clip,scale=0.7]{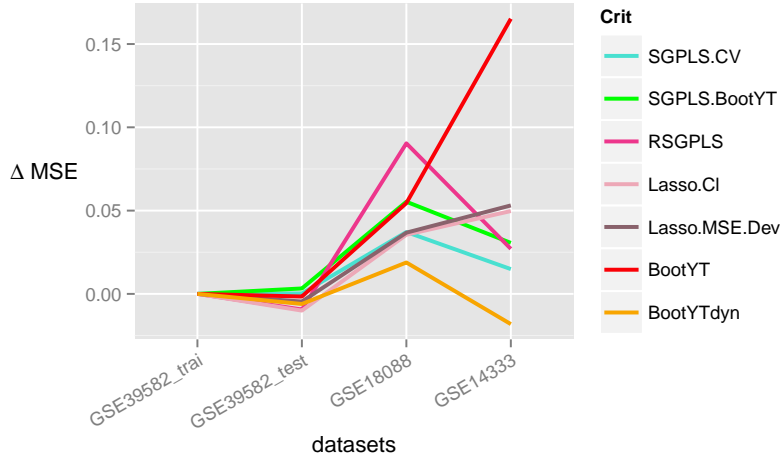}
	\caption{Differences between the MSE computed on the training subset of the original dataset and the PMSE obtained on test sets.}
	\label{fig.6.11}
\end{figure}

In this real data study,  BootYTdyn retains only one predictor, which is also retained by all other methods. Thus, as expected, this most sparse support induced the highest values of both the MSE and number of misclassified observations, based on the training subset of the original dataset. It also provided the highest values based on the test subset of the original dataset, which is to be expected since, as explained in Section \ref{321}, these two parts are well-balanced in terms of anatomo-clinical characteristics. 
Thus, this causes a bias in evaluation of model's predictive abilities, by making comparable MSE and misclassified values based on both the training and testing subsets. Results in Table~\ref{tab.6.12} confirm this property, and also highlight the usefulness of additional independent  datasets for reliably comparing predictive abilities. While a well-designed model for predictive purposes will provide similar PMSE values on comparable independent additional datasets compared to MSE obtained on the training dataset, an over-fitted one would be related to higher PMSE values, due to its dependance on training data. Thus, the differences, noted $\Delta$ MSE, between the MSE obtained on the training subset of the original dataset, and those obtained on the three test datasets, are shown in Fig.~\ref{fig.6.11}.

Our new dynamic bootstrap-based approach is the only one that exhibits this MSE stability property, while all others provided higher PMSE values on the two additional datasets than on the original one. This led  BootYTdyn to have only 48 (16\%) misclassified values on these two additional datasets, which represents the best result among all studied approaches. Thus, we can reasonably assume that our new method has helped to remove non-informative predictors, in the sense of not being relevant for improving predictive ability.

Lastly, the unique extracted probe set, named 230784\_at, is already known to be related to the original location in the distal colon. The sign of the regression coefficient obtained with  BootYTdyn  is coherent with this state-of-the-art result, and thus strengthens our conclusion.
  
\section{Discussion}
\label{5}

In this article, we developed a new dynamic bootstrap-based technique for variable selection, suitable for both the PLS and GPLS frameworks, and proposed a bootstrap-based version of the SPLS and SGPLS methods for selecting the number of components. While the first of these lets us completely avoid CV, the second  lets us select the set of tuning parameters in a more reliable way. 

In state-of-the-art approaches, the use of CV-based techniques for selection of hyper-parameters is common, and can lead to important stability issues, observed notably by \citet{BoulesteixCV14} and \citet{magnanensibootcrit}, and confirmed in our studies. \citet{sun2013consistent} also worked on this subject, and proposed a methodology for selecting  tuning parameters of penalized regressions in order to stabilize variable selection. Even if their method is not applicable to the selection of the number of components in PLS regression, it would be interesting to adapt it to both SPLS CV and our new SPLS BootYT implementation, in order to look for potential stability gains. However, our new dynamic bootstrap-based technique represents a useful method for avoiding CV-related issues, since the unique hyper-parameter is successively selected using a bootstrap-based criterion. This new technique improves on the original bootstrap-based methodology introduced by \citet{lazraq2003selecting}, in that it permits us to approximate the distribution of covariates' regression coefficients by removing the condition of working in a  subspace of fixed dimension $K$. Theoretical results have been established that strengthen the usefulness of building subspaces spanned by a dynamic number of components for performing PLS regressions on  bootstrap samples.    

In the PLS framework, conclusions based on our simulations are twofold. First, for datasets with negligible random noise in $\mathbf{y}$,  SPLS BootYT is recommended. Indeed, both in terms of accuracy and predictive abilities, it outperforms all  other techniques compared here. Second, for datasets with non-negligible random noise in $\mathbf{y}$, which represents the more realistic case, our new dynamic bootstrap-based method is recommended. As for  SPLS BootYT for the negligible random noise variability framework,  BootYTdyn outperform all other methods for each property studied. Furthermore, it is the only method which concludes as to a decreasing number of significant predictors, while the random noise variability increases, which was expected here.

Results obtained from our classification study using real datasets match with previous conclusions. Indeed,  BootYTdyn is the only one which leads to the expected PMSE values on two additional independent datasets, which allows us to suppose that over-fitting issues were avoided, and that these PMSE are induced by noise or information which cannot be modeled using only gene expression. Furthermore, the extracted probe set is already known to be linked to the relevant location in the distal colon, which strengthens our confidence in this new dynamic approach.

Lastly, our new bootstrap-based SPLS implementation improves  the stability of this method. Indeed, in all cases studied, both the SPLS CV and SGPLS CV conclude in a higher number of different sets of hyper-parameters than our bootstrap-based versions do, leading to higher numbers of different supports and models too.

In the future, simulations need to be done to booster the results obtained on real datasets for the logistic framework (Section \ref{4}). Testing the performance of these new approaches for responses following other distributions also must be done. However, based on all results obtained in these studies, our new dynamic method appears to  be the most efficient compared to state-of-the-art approaches for datasets with non-negligible noise variability, a common situation  in daily practice. 

\section*{Supplementary material}

\begin{description}

\descitem{Theoretical results.} Proofs of several theoretical results related to the number of components in bootstrap samples extracted from an original dataset. (Math\_nb\_comp\_ boot.pdf)

\descitem{Predictors' correlations properties.} A summary of predictors' correlations related to simulations in Section \ref{32}. (Resume\_corr\_pred.pdf)

\descitem{Numbers of extracted components.} A summary of the extracted numbers of components used for performing the original bootstrap-based method on simulated datasets in section \ref{32}. (Nb\_comp\_sim.pdf)

\end{description}

\section*{Acknowledgements}

\textit{Funding}: The authors gratefully acknowledge the Labex IRMIA for J. Magnanensi's PhD grant.

\bibliographystyle{natbib}
\bibliography{biblio}
\end{document}